\documentclass[aps,superscriptaddress,eqsecnum,nofootinbib,showpacs,preprintnumbers]{revtex4-2}
\usepackage{graphicx}
\usepackage{epsfig}
\usepackage{bm}
\usepackage{amssymb}
\usepackage{float}
\usepackage{amsmath}
\usepackage{dcolumn}
\usepackage{cancel}

\usepackage{mathrsfs}
\usepackage{dcolumn}
\usepackage{graphicx}
\usepackage{amsmath}
\usepackage{amsfonts}
\usepackage{amssymb}
\usepackage{subfigure}
\usepackage{makeidx}
\usepackage{bm}
\usepackage{epsf}
\usepackage{color}
\usepackage{multirow,dcolumn}
\usepackage{graphicx}
\usepackage{mathrsfs}
\graphicspath{{Images/}}
\newcommand{\cb}{\color{black}}
 \newcommand{\cbl}{\color{black}}

\def\doi{http://doi.org}

\def\be{\begin{equation*}}
\def\ee{\end{equation*}}

\newcommand{\chiT}{\chi^{\mu\nu}{}_{\rho\sigma}}
\newcommand{\Qeff}{Q^{\rm eff}_e}

\begin{document}

\preprint{\leftline{KCL-PH-TH/2026-{\bf 14}}}

\title{Self-Gravitating Magnetic Monopoles and Dyons in String-Inspired Models:\\Structure and Stability}

\author{Nick E. Mavromatos}
\email{nikolaos.mavromatos@kcl.ac.uk}
 \affiliation{Physics Division, School of Applied Mathematical and Physical Sciences, National Technical University of Athens, 15780 Zografou Campus,
Athens, Greece.}

\affiliation{Theoretical Particle Physics and Cosmology Group, Department of Physics, King's College London, London WC2R 2LS, United Kingdom.}

\author{Sarben Sarkar}
\email{sarben.sarkar@kcl.ac.uk}
\affiliation{Theoretical Particle Physics and Cosmology Group, Department of Physics, King's College London, London WC2R 2LS, United Kingdom.}

\author{Dionysios P. Theodosopoulos}
	\email{d.theodosopoulos@utexas.edu}
	\affiliation{Texas Center for Cosmology and Astroparticle Physics, Weinberg Institute for Theoretical Physics, Department of Physics, The University of Texas at Austin, Austin, TX 78712, USA}

\vspace{4.5cm}

\begin{abstract}
\cb

We present classical magnetic--monopole (MM) and dyon solutions induced by global
monopoles in string-inspired gravitational models. Although several ingredients
of the construction have appeared previously, we examine a new synthesis via
their organised use in a self-gravitating finite-energy monopole/dyon system
with explicit mass, charge, core and stability diagnostics. The global monopole
provides the topological seed, while the dilaton, Kalb--Ramond (KR) axion and
Born--Infeld (BI) electromagnetic (EM) sector supply the string-inspired dressing. Two related branches are analysed. The first is a dilaton--BI
magnetic branch, distinct from earlier KR--induced MM 
constructions; the second, is a dyonic extension in which the KR axion couples electric and magnetic sectors through an axion-Pontryagin-anomaly interaction.
In both cases BI dynamics regularises
the EM self-energy within a self-gravitating global-monopole
construction, and the solutions possess a de Sitter core, a positive ADM-mass
parameter range and satisfy the standard energy conditions. A further novel 
feature is a limiting branch in which the ADM mass tends to zero while the
magnetic charge remains finite, the configuration being supported by magnetic
charge and de Sitter core pressure.We also examine stability diagnostics. The stress--energy tensor gives
finite force components and an outward-pointing radial force, indicating
absence of core collapse within the mechanical stability criterion. In the exterior
analytic region we analyse gauge-invariant EM perturbations
using the Gervalle-Volkov framework. The helicity modes $\psi_\pm$ obey a
self-adjoint Sturm--Liouville system. MM helicities decouple,
while dyons exhibit axion-induced helicity mixing and birefringent
polarisation structure. Positivity of the EM helicity block
provides a necessary long-range stability diagnostic; the full coupled
Einstein-matter spectral problem is discussed.

\end{abstract}

\maketitle

\flushbottom


\section{Introduction}

Although magnetic monopoles (MM) are theoretical solutions of the Euler-Lagrange equations of some microscopic field-theory models, there is still no experimental evidence~\cite{Mavromatos:2020gwk} for their existence, despite intensive searches, in both the Cosmos and at colliders.
There may be various reasons for this. 
One could be that such configurations arise in Grand Unified Theory (GUT) models~\cite{tHooft:1974kcl,Polyakov:1974ek}, in which the electromagnetic U(1) gauge group is part of a larger group, {\it e.g.} a GUT gauge group, or is embedded in an SU(2) gauge group in Georgi-Glashow-type field theories~\cite{Georgi:1974sy}.
In the GUT case, the MM solutions
have masses close to the inflationary scale (that is, close to grand unification scale (GUT), $M \gtrsim 10^{14}~{\rm GeV}$). Such MM have been diluted by inflation, and, thus, their observation is practically impossible. Another reason for the non-observation of such MM, even if they had 
relatively low masses (of the order of the electroweak scale), is their compositeness. Such (solitonic) MM solutions consist of gauge-bosons and (charged) Higgs quanta; their production in current or future colliders is exponentially suppressed~\cite{Drukier:1981fq}, despite their theoretical justification as being solutions of field equations in specific models~\cite{Cho:1996qd,Cho:2013vba,Ellis:2016glu} 
(based on the Standard-Model (SM) gauge group, or appropriate extensions~\cite{Kephart:2017esj}). 
In collider kinematics, a composite monopole, such as a
\textit{'t Hooft--Polyakov} or \textit{Cho--Maison} monopole, is not
created as an elementary point particle.  Rather, it would have to be
assembled from many Higgs and gauge-field quanta into a coherent
classical soliton.  This produces an entropy-mismatch penalty: among all
possible high-multiplicity final states with the same energy, only an
exponentially small fraction have the ordered field configuration of a
monopole.  The corresponding rate is therefore expected to be suppressed
by a factor of order
\[
  \exp\!\left(-\frac{4}{\alpha}\right)
  \sim 10^{-240},
\]
where $\alpha\simeq 1/137$ is the fine-structure constant.  Collider
production of such composite monopoles is consequently unobservable,
even when their masses are kinematically accessible.

The situation is radically different in the early Universe. The
entropy-suppression argument applies to collider production because a few
incoming quanta must organise themselves into a coherent, extended soliton.
During a cosmological symmetry-breaking phase transition, however, the
relevant Higgs field is already a macroscopic classical field distributed
throughout space. Monopoles are then formed by the ordering of the vacuum
itself, rather than by the rare assembly of many particles into a special
final state.
Suppose a symmetry group $G$ is spontaneously broken to a subgroup $H$. The
degenerate vacuum states form the vacuum manifold
\begin{equation}
  \mathcal{M}_{\rm vac} = \frac{G}{H}.
\end{equation}
At temperatures above the critical temperature $T_c$, the symmetry is
restored and the Higgs expectation value vanishes. As the Universe cools
through $T_c$, the Higgs field rolls away from the symmetric point and
chooses a vacuum value on $\mathcal{M}_{\rm vac}$. The crucial point is causality. At the time of the transition, two regions
separated by more than a horizon distance cannot communicate. Therefore they
cannot coordinate their choices of Higgs orientation. The Higgs field chooses
its vacuum direction independently in different causal domains. If the
correlation length is denoted by $\xi$, then one expects roughly independent
vacuum choices over regions of size $\xi$, with
\begin{equation}
  \xi \lesssim H^{-1}(T_c),
\end{equation}
where $H(T_c)$ is the Hubble rate at the transition. For monopoles the relevant topology is
\begin{equation}
  \pi_2(\mathcal{M}_{\rm vac}) \neq 0.
\end{equation}
In the simplest case,
\begin{equation}
  \pi_2(\mathcal{M}_{\rm vac}) = \mathbb{Z}.
\end{equation}
This means that maps from a two-sphere in physical space into the vacuum
manifold can carry an integer winding number. Around a would-be monopole
core, the Higgs field defines a map
\begin{equation}
  S^2_{\rm space} \longrightarrow \mathcal{M}_{\rm vac}.
\end{equation}
If this map has non-zero winding number, the Higgs field cannot be smoothly
deformed everywhere into a single vacuum orientation. At the centre of the
configuration the Higgs field must leave the vacuum manifold, typically
passing through $\phi = 0$. That point is the monopole core.
 
This is the Kibble mechanism~\cite{Kibble:1976sj}. Random choices of vacuum
orientation in neighbouring domains make it unavoidable that some closed
surfaces in space enclose non-trivial winding. The monopole is therefore not
produced by a rare collision process; it is a topological remnant of the
phase transition.
A crude estimate of the initial monopole density follows immediately. If
approximately one independent field orientation is chosen per correlation
volume $\xi^3$, then the initial number density is of order
\begin{equation}
  n_M(t_c) \sim \xi^{-3}.
\end{equation}
Since causality gives $\xi \lesssim H^{-1}(T_c)$, one often writes the rough
Kibble estimate as
\begin{equation}
  n_M(t_c) \sim H^3(T_c),
\end{equation}
up to numerical factors and model-dependent corrections. Subsequent
monopole--antimonopole annihilation, entropy production, inflationary
dilution, or plasma effects can reduce this abundance, but the important
point is that the initial formation probability is not Boltzmann suppressed.
 Thus the early-Universe phase transition avoids the collider
entropy-mismatch penalty. A composite monopole does not have to be assembled
from many uncorrelated quanta. Instead, it forms as a classical field defect
because the Higgs vacuum orientation cannot be chosen consistently across all
causally disconnected domains.

Hence the unsuppressed channels for producing composite magnetic
monopoles are (i) symmetry-breaking phase transitions in the early
Universe and (ii) the dual Schwinger mechanism in ultra-strong magnetic
fields, both of which circumvent the exponential suppression factor
characteristic of collider environments~\cite{Affleck:1981ag,Gould:2017zwi,Gould:2019myj,MoEDAL:2021vix}, which catalyze the unsuppressed production of MM. This suppression does not apply to elementary (structureless) Dirac-type MM~\cite{Dirac:1948um,Dirac:1931kp}, behaving as sources of singular magnetic fields (magnetic poles), although their mass needs to be made finite by mechanism which go beyond the standard model \cite{Mavromatos:2026nlp}. We characterise these as ordinary monopoles \cite{Shnir:2005vvi} since:
\begin{itemize}
  \item They are \emph{point-like}: the monopole magnetic field is exactly that of
        an isolated electric charge but with roles of $\mathbf E$ and
        $\mathbf B$ swapped,
        \[
          \mathbf B(\mathbf r)=\frac{g}{4\pi}\,\frac{\mathbf r}{r^{3}},
          \qquad g = n\,g_{\rm D},\; n\in\mathbb Z .
        \]
  \item Dirac’s quantisation condition,
        $e\,g = \tfrac{n}{2}\,\hbar c$, fixes the fundamental charge
        $g_{\rm D}$ and ensures single-valued electron wavefunctions.
  \item It carries \emph{no internal structure}: there is no Higgs
        field or non-Abelian gauge profile inside a finite radius.
\end{itemize}
The monopoles that we will construct will approximate such monopoles at late epochs (once the relevant dilaton time gradients fade after the formation epoch, right after the symmetry breaking phase transition) \footnote{Although the self-gravitating Born–Infeld monopole has a finite-radius
core and scalar hair, its \emph{external} electromagnetic field is
identical to that of a pointlike Dirac monopole.}.

 Since gravity can affect their masses, because of gravitational binding energy, MM solutions need, for completeness, to incorporate gravity. The masses compared to the Minkowski spacetime cases, could be reduced. This is necessary for experimental verification, and also  essential when considering MM in the early Universe. 
Self-gravitating magnetic monopoles in string-inspired effective
theories provide a natural setting in which nontrivial topology,
nonlinear gauge dynamics, and spacetime backreaction are realised within
a single framework. The inclusion of dilaton and
Born--Infeld sectors, motivated by low-energy string theory~\cite{Fradkin:1985qd}, can lead to
finite-energy monopole solutions with regular cores and nontrivial
asymptotic structure~\cite{Mavromatos:2018kcd}. The Born--Infeld (BI) form of nonlinear electrodynamics is
singled out both by its string-theoretic origin and by its regularisation of the 
divergent electromagnetic self-energy of a monopole. This allows solutions with a
finite, positive ADM mass (the total gravitational energy seen at infinity).  
Moreover, as discussed in \cite{Mavromatos:2018kcd}, Born-Infeld regularised theories can also admit structureless Dirac-type MM solutions, which can be produced at colliders without suppression, as mentioned above.

It is our purpose  to discuss such solutions, which stem from non-gauged global monopoles, but are promoted to MM through appropriate mediator fields, for example scalar dilatons or pseudoscalar axions, that characterise microscopic string models~\cite{Svrcek:2006yi}. One such case~\cite{Mavromatos:2016mnj} arose in string-inspired models in which a MM arises 
from global monopoles~\cite{Barriola:1989hx}, as a result of the coupling of a (constant, but non-trivial) dilaton scalar field to both the electromagnetic Maxwell part of the effective action and the kinetic terms of the Kalb-Ramond (KR) field strength of the spin-one massless antisymmetric tensor of the massless bosonic string gravitational multiplet~\cite{Green:2012oqa,Green:2012pqa}. In (3+1) spacetime dimensions, after string compactification, the KR field is dual to a massless pseudoscalar (KR axion) field, and thus the induced MM has an axion charge, that is quantized in order for the Dirac charge quantization condition~\cite{Dirac:1931kp} to be satisfied. The gravitational stability and finiteness of total energy for such MM have been discussed~\cite{Mavromatos:2018drr}, while their sizes and masses have been estimated, in terms of the parameters of the model.

In this work we  treat other interesting cases in effective string-inspired actions containg gravity, where global monopoles with dilaton or axion (KR) charges  are promoted to MM via appropriate interactions.~Two families of such solutions are 
constructed (partly numerically) within an action containing gravity, a global-O(3) Higgs
triplet, a dilaton $\Phi$, an axion (KR pseudoscalar) $b$, and Born--Infeld
electrodynamics~\cite{Fradkin:1985qd}.  The first family consists of purely magnetic monopoles stabilised by a \emph{non-constant}
dilaton with charge $\zeta$;\footnote{The dilaton charge is the coefficient of the $1/r$  \emph{fall-off} of the dilaton field at large radial distance.} the second extends these to dyons when the axion field has a  $bF\tilde{F}$ coupling (where $F$ denotes the electromagnetic field strength and $\tilde{F}$ its dual).
Both families are shown to possess positive ADM mass and a regular de Sitter core,
with the form of the metric, dilaton, and Higgs profiles verified partly numerically. 

\cb

The novelty of the present work is not the separate use of global monopoles,
Kalb--Ramond axions or Born--Infeld electrodynamics, all of which have
appeared previously, but their organised use in a self-gravitating
finite-energy monopole/dyon construction. We identify a
dilaton--Born--Infeld magnetic branch, distinct from earlier
Kalb--Ramond-induced monopole constructions, in which the magnetic charge is
generated by the coupling of a non-constant massless dilaton to the
Born--Infeld electromagnetic sector and satisfies
\begin{equation}
  Q_m^2 = M\zeta,
\end{equation}
with $\zeta$ the dilaton scalar charge and $M$ the ADM mass. We then
construct the corresponding dyonic extension, in which the Kalb--Ramond axion
couples electric and magnetic sectors through
$b\,F_{\mu\nu}\widetilde{F}^{\mu\nu}$, leading to the unified charge
relation
\begin{equation}
  Q^2 = Q_e^2 + Q_m^2 = M\zeta.
\end{equation}
For both branches we obtain a self-gravitating core--exterior description
with a regular de~Sitter core, identify the positive ADM-mass parameter
range, and verify the finite Born--Infeld electromagnetic self-energy. We
also find a limiting branch in which the ADM mass tends to zero while the
magnetic charge remains finite, the configuration being supported by magnetic
charge and de~Sitter core pressure. 
\cbl

A central question for such configurations is their stability, both mechanical in the sense discussed in \cite{Farakos:2025byy}, and 
dynamical. 
While the existence of regular solutions is by now established,
their mechanical stability viewed as solitonic objects  (similar to configurations appearing in nuclear matter~\cite{Laue:1911lrk}) needs to be examined first. If
the total internal force is finite and its radial component is directed inwards, towards the centre of the monopole, then this would imply the tendency of the system to collapse leading to mechanical instability. By contrast a finite total force, radially pointing outwards,  
 would indicate mechanical stability. Moreover, stability under linear perturbations of such solutions needs also to be studied, and provides important complementary  information.\footnote{Such linear-stability analyses is a standard tool in the case of solitonic field-theoretic solutions in curved spacetimes, \emph{e.g.} black holes with Higgs hair in spontaneously broken SU(2) gauge theories~\cite{Mavromatos:1995kc}. Needless to say, though, that linear stability by itself does not provide a proof of absolute stability, and must be 
 extended to the case beyond linear perturbations, which algebraicailly is often a very complicated task, or combined with other methods to get complementary information (\emph{e.g.} Catastrophy theory, as in the case of \cite{Mavromatos:1995kc}).} This task 
is complicated in our case, owing to
nonlinear constitutive relations and the coupling between the gauge,
scalar, and gravitational sectors.

In contrast with simple single-field stability problems, the
Born--Infeld--dilaton--axion monopole does not generally lead to a single
elementary effective potential $V(r)$. In the simplest case one would obtain
a Schr\"{o}dinger-type equation of the form \footnote{Here $r_*$ is the toroise co-ordinate used in black hole physics.}
\begin{equation}
  -\frac{d^2 u}{dr_*^2} + V(r)\,u = \omega^2 u,
\end{equation}
where $V(r)$ is a known scalar function. If $V(r)$ is manifestly non-negative,
stability can often be inferred directly.
In the present problem the situation is more complicated. The static, spherically symmetric metric is written as
\begin{equation}
  ds^2
  =
  -B(r)\,dt^2
  +
  \frac{dr^2}{A(r)}
  +
  R^2(r)d\Omega^2 .
  \label{eq:background_metric_radial_functions}
\end{equation}
and the background
contains several non-trivial radial functions,
\begin{equation}
  A(r), \qquad B(r), \qquad R(r), \qquad \Phi(r), \qquad b(r),
  \qquad F_{\mu\nu}(r),
\end{equation}
The remaining radial functions are matter fields.  The dilaton profile
\begin{equation}
  \Phi=\Phi(r)
\end{equation}
changes the effective electromagnetic coupling, since the Born--Infeld
sector contains factors such as
\begin{equation}
  e^{\Phi},
  \qquad
  e^{-2\Phi}.
\end{equation}Therefore the electromagnetic perturbations propagate in a medium whose
effective coupling varies with radius. In the dyonic case there is also a Kalb--Ramond axion profile
\begin{equation}
  b=b(r),
\end{equation}
which couples through the pseudoscalar term
\begin{equation}
  b\,F_{\mu\nu}\widetilde F^{\mu\nu}.
\end{equation}
Because $b(r)$ is radius-dependent, it induces radius-dependent mixing
between electric and magnetic perturbations, or equivalently between the
two helicity sectors. Finally,
\begin{equation}
  F_{\mu\nu}=F_{\mu\nu}(r)
\end{equation}
denotes the background electromagnetic field strength.  In the purely
magnetic case its non-zero component is
\begin{equation}
  F_{\theta\phi}=Q_m\sin\theta ,
\end{equation}
whereas in the dyonic case there is also an electric component,\begin{equation}
  F_{tr}(r)
  =
  e^{\Phi(r)}
  \frac{Q_e-\bigl(b(r)-b_0\bigr)Q_m}{R^2(r)} .
  \label{eq:dyonic_electric_component_explanation}
\end{equation}
Consequently, the perturbation equations have coefficients depending on
\begin{equation}
  A(r),\qquad B(r),\qquad R(r),\qquad
  \Phi(r),\qquad b(r),\qquad F_{\mu\nu}(r).
\end{equation}
They are therefore not governed by a simple potential such as
\begin{equation}
  V(r)=\frac{\ell(\ell+1)}{r^2}.
\end{equation}
The electromagnetic perturbations also probe the Born--Infeld constitutive
tensor. The latter depends nonlinearly on the background field invariants
\begin{equation}
  F_{\mu\nu}F^{\mu\nu}, \qquad F_{\mu\nu}\widetilde{F}^{\mu\nu}.
\end{equation}
Consequently, after harmonic decomposition, the perturbation equations do not
reduce to a single closed-form potential. Instead, they take the schematic
Sturm--Liouville form
\begin{equation}
  \mathcal{L}\,\Psi
  = -\frac{d}{dr_*}\!\left(P(r)\frac{d\Psi}{dr_*}\right)
    + \mathbf{V}(r)\,\Psi,
  \label{eq:SL_structural_form}
\end{equation}
where $\Psi$ denotes the vector of radial gauge-invariant perturbation amplitudes and $\mathbf{V}$ is a matrix.The tortoise coordinate $r_*$ is defined by
\begin{equation}
  \frac{dr_*}{dr}
  =
  \frac{1}{\sqrt{A(r)\,B(r)}}.
  \label{eq:tortoise_def}
\end{equation}
Equivalently,
\begin{equation}
  r_*(r)
  =
  \int^r
  \frac{d\bar{r}}{\sqrt{A(\bar{r})\,B(\bar{r})}}.
  \label{eq:tortoise_integral}
\end{equation}

In particular, the perturbation
equations do not, in general, reduce to a simple closed-form potential,
and must instead be analysed through their structural properties.
Specifically, we consider \emph{dynamical stability} of electromagnetic perturbations of a
self-gravitating monopole background and show that the problem reduces
to a self-adjoint Sturm--Liouville system of the above type. For the purely magnetic
monopole, the helicity sectors decouple identically, and the analysis
reduces to a scalar radial equation defined on a piecewise background
consisting of an \emph{interior} de~Sitter core and an {exterior} monopole
geometry.  For gravitating solitons this analysis
is non-trivial because gauge redundancy, constraint equations, and mode mixing of different helicities in
curved backgrounds all conspire to obscure the physical degrees of freedom if one were to work
with the perturbation of the vector potential.

Although the effective radial potential in the Sturm-Liouville equation cannot be written in  closed analytic form, its qualitative structure is sufficient to
control the spectrum. In particular, the interior region is governed by
a positive centrifugal barrier, while the exterior potential exhibits a
repulsive divergence near the inner boundary and a non-negative
asymptotic tail. We formulate the stability problem in variational
terms and show that, under mild and physically well-motivated
assumptions on  subleading contributions from the Born--Infeld and
dilaton sectors, no negative modes arise.
The systematic framework for addressing this question is due to
Gervalle-Volkov (GV) ~\cite{Gervalle:2022npx}, and was developed originally for
perturbations of electroweak monopoles (Cho-Maison (CM))~\cite{Cho:1996qd}. Recently, this method has been applied~\cite{Mavromatos:2026nlp} also to study the stability of the BI regularised CM~\cite{Arunasalam:2017eyu,Mavromatos:2018kcd}. We note that stringent lower bounds on the mass of such BI monopoles can be imposed at current and future colliders~\cite{Ellis:2017edi,Ellis:2022uxv,Mitsou:2026gvp}.
The overlap with the GV framework is methodological rather
than physical: as in GV and in the BI-regularised electroweak monopole
analysis of \cite{Mavromatos:2026nlp}, we use  complex-tetrads, spin-weighted
harmonics, gauge-fixing, and constraint-elimination machinery to reduce
the perturbation problem to a self-adjoint radial Sturm-Liouville
system. The novelty here is its application  to a
self-gravitating dilaton-axion Born--Infeld monopole/dyon background,
where the constitutive tensor induces a helicity-based radial system and,
in the dyonic case, a nontrivial mixing term $W(r)$.  The GV approach works directly with
gauge-invariant variables built from the field-strength perturbation $\delta F_{\mu\nu}$,
projects these onto helicity eigenstates using a null tetrad, and exploits the constitutive
tensor $\chi^{\mu\nu}{}_{\rho\sigma}$ of nonlinear electrodynamics to reduce the
linearised stability equations to a Schr\"{o}dinger-like radial system of Sturm-Liouville type.
For static backgrounds the resulting operator is self-adjoint, immediately implying a
real spectrum and hence stability.

The present paper provides a self-contained implementation of the GV analysis for our
solutions.  Our aim is threefold: (i) to derive the gauge-invariant helicity variables
from first principles in an  accessible manner; (ii) to reduce the linearised equations explicitly to the Sturm--Liouville form
and determine the effective potentials; (iii) to use the structure of the resulting operator
to establish linear stability for both monopoles and dyons. For completeness, we also check that the energy conditions of such solutions are satisfied. \cb Positivity of the
electromagnetic helicity block provides a necessary long-range stability
diagnostic. 

\color{black} The present analysis follows the GV construction for the electromagnetic sector. Unlike the electroweak Cho–Maison monopole~\cite{Cho:1996qd}, however, the present theory contains additional physical scalar degrees of freedom—the dilaton, Kalb–Ramond axion and global-monopole Higgs—which cannot be removed by gauge fixing. A complete Einstein–matter reduction analogous to GV, therefore, constitutes a substantially larger problem and lies beyond the scope of the present work. \cbl

The paper is organised as follows.  Section~\ref{sec:GM} gives a brief review of the self-gravitating global monopole of \cite{Barriola:1989hx}. Section~\ref{sec:GlobalDilaton} discusses magnetic monopoles induced by global monopoles with a dilaton charge, and Section~\ref{sec:GlobalAxion} treats the dyonic generalisation with dilaton and axion charges in models incorporating Born--Infeld electrodynamics. Section~\ref{sec:stab} contains the full stability analysis. We first establish mechanical stability and the energy conditions (Section~\ref{sec:mechstab}). Section~\ref{sec:stabfoprm} then develops the formalism for piecewise linear dynamical stability, introducing gauge-invariant perturbation variables, the helicity projections $\psi_\pm$, and the self-adjoint Sturm--Liouville structure of the perturbation equations. Section~\ref{sec:stabanal} applies this framework to the explicit solutions: Section~\ref{sec:monopole} treats the purely magnetic monopole (where the helicity sectors decouple and the effective potential is positive definite) and Section~\ref{sec:dyon} treats the dyonic case, deriving the helicity-mixing term and establishing stability from self-adjointness and an explicit bound on the off-diagonal potential. Finally, Section~\ref{sec:conclude} summarises our conclusions. Appendix~\ref{app:projection_monopole_full} gives the detailed derivation of the radial potentials for the monopole case, 
\color{black} whilst Appendix \ref{subsec:epsilon_mixing} discusses briefly the full coupled Einstein--matter spectral problem, whose detailed study is left
for future work, due to its algebraic complexity. \color{black} Appendix~\ref{app:mono_mech_stab} provides supplementary material on mechanical stability.

\section{Global Monopoles}\label{sec:GM}

Global monopoles (GMs) are topological defects arising from the spontaneous breaking of a global O(3) symmetry. The gravitational field of a GM in Einstein gravity was first derived in Ref.~\cite{Barriola:1989hx}, considering a Higgs triplet $\chi^{a}$ (with $a=1,2,3$) and a Higgs potential that breaks the global O(3) symmetry:
\begin{equation}
    S=\int d^{4}x\sqrt{-g}\left[\frac{R}{16\pi G}+\mathcal{L}_{GM}\right]~,
\end{equation}
\begin{equation}
    \mathcal{L}_{\text{GM}}=-\frac{1}{2}(\nabla_{\mu}\chi^{a})(\nabla^{\mu}\chi^{a})-\frac{\lambda}{4}\left(\chi^{a}\chi^{a}-\eta^{2}\right)^{2}~,\label{LGM}
\end{equation}
where $g=det\left(g_{\mu\nu}\right)$, and $R$ is the Ricci scalar~\footnote{Our conventions and definitions throughout this work are: $(-,+,+,+)$ for the signature of the metric, the Riemann tensor is defined as 
$R^\lambda_{\,\,\,\,\mu \nu \sigma} = \partial_\nu \, \Gamma^\lambda_{\,\,\mu\sigma} + \Gamma^\rho_{\,\, \mu\sigma} \, \Gamma^\lambda_{\,\, \rho\nu} - (\nu \leftrightarrow \sigma)$, 
and the Ricci tensor and scalar are given by  $R_{\nu\alpha} = R^\lambda_{\,\,\,\,\nu \lambda \alpha}$ and $R= g^{\mu\nu}\, R_{\mu\nu}$ respectively. 
Also we work in natural units $\hbar=c=1$. Newton's
constant $G$ is retained explicitly.}. In the region where the O(3) symmetry is spontaneously broken, the spacetime metric reads
\begin{equation}
    ds^{2}=-\left(1-8\pi G \eta^2-\frac{2GM}{r}\right)dt^{2}+\frac{dr^{2}}{1-8\pi G \eta^2-\frac{2GM}{r}}+r^{2}\left(d\theta^{2}+\sin^{2}\theta d\phi^{2}\right)~,\label{GM}
\end{equation}
where $\eta$ determines a deficit solid angle. Specifically, after appropriate coordinate redefinitions, the spacetime \eqref{GM} exhibits a conical singularity associated with the angular deficit factor $1-8\pi G\eta^2$~\cite{Barriola:1989hx,Mavromatos:2016mnj}.

A physically acceptable GM spacetime should be free of horizons and curvature singularities. In that sense, the authors of Ref.~\cite{Harari:1990cz} proposed that the monopole possesses a de Sitter (dS) core of radius $\delta$, within which the metric is given by
\begin{equation}
    ds^{2}=-\left(1-\frac{2\pi G\lambda\eta^4}{3}r^2\right)dt^{2}+\frac{dr^{2}}{1-\frac{2\pi G\lambda\eta^4}{3}r^2}+r^{2}\left(d\theta^{2}+\sin^{2}\theta d\phi^{2}\right)~.\label{metricglobal}
\end{equation}

Imposing the Israel conditions~\cite{Israel:1966rt}, which require continuity of the metric and its first derivative across the core boundary, yields the ADM mass and core radius
\begin{equation}
    M=-\frac{16\pi}{3}\frac{\eta}{\sqrt{\lambda}}, ~ \delta=\frac{2}{\eta\sqrt{\lambda}}~.
\end{equation}
The negative ADM mass implies repulsive gravitational effects, rendering the GM configuration unstable. This result was further confirmed numerically in Ref.~\cite{Harari:1990cz}, where the mass was found to be
\begin{equation}
    M=-6\pi\frac{\eta}{\sqrt{\lambda}}~,
\end{equation}
indicating that the instability persists beyond the analytic approximation. This motivates the search for extended frameworks in which the monopole mass can become positive, as we explore in the following sections.

\section{Magnetic Monopoles from Global Monopoles with a Dilaton Charge}\label{sec:GlobalDilaton}

In this section we construct a magnetically charged monopole solution generated by a GM coupled to gravity, in the presence of a dilaton field and Born–Infeld non-linear electrodynamics. Our motivation is twofold. First, previous studies have shown that suitable extensions of the gravitational sector can convert the negative-mass behaviour of ordinary GMs into physically acceptable configurations with positive ADM mass. Second, in string-inspired settings the dilaton naturally couples to the electromagnetic (EM) sector, while the Born–Infeld action provides a non-linear completion of Maxwell theory that regularises the monopole self-energy. Against this background, we investigate whether a GM can source a magnetic charge through its coupling to the dilaton and thereby give rise to a regular, self-gravitating magnetic monopole with positive ADM mass. We first formulate the Einstein-frame field equations, then derive approximate analytic solutions inside and outside the monopole core, and finally compare them with numerical results.

A GM solution with positive ADM mass in an extended Gauss–Bonnet theory of gravity was derived in \cite{Chatzifotis:2022ubq}. Moreover, a magnetically charged GM solution with positive ADM mass was obtained in a string-inspired gravitational theory \cite{Mavromatos:2016mnj,Mavromatos:2018drr}, where the magnetic charge originates from a Kalb–Ramond axion-like field and ensures the positivity of the ADM mass. Motivated by this result, we seek a magnetic monopole solution sourced by a GM minimally coupled to gravity, in which the magnetic charge is associated with a dilaton charge and leads to a positive ADM mass. The EM sector of the Lagrangian will be described by the string inspired Born-Infeld (BI) electrodynamics. 

One of the distinctive features of string/brane–induced non-linear electrodynamics is that higher-order corrections in the Maxwell tensor can be resummed into a closed-form expression, namely the BI Lagrangian \cite{Born:1933qff,Born:1934ji,Born:1934gh,Metsaev:1987qp,Andreev:1988cb,Tseytlin:1999dj}. This arises from the resummation of open-string excitations, for example those attached to D3-brane worlds in the D-brane formulation of string theory. In this case, the three-dimensional brane world-volume leads to the Dirac–Born–Infeld (DBI) action (see \cite{Leigh:1989jq,Dai:1989ua,Polchinski:1996na} and references therein). In such constructions, BI electrodynamics in four spacetime dimensions originates from the higher-dimensional ($d=10$) superstring action, either through compactification or by restriction to the world-volume of a D3-brane. Importantly, in all string-inspired scenarios the BI Lagrangian couples to the inverse of the open-string coupling, $g_s = e^{\phi}$, where $\phi$ denotes the dimensionless dilaton field. Consequently, the effective four-dimensional action in a curved background metric (in the Jordan or $\sigma$-model frame) $g^{J}_{\mu\nu}$ takes the form
\begin{equation}
S^{J}_{\text{BI}} = -T_{4}^{2} \int d^{4}x \, e^{-\phi} 
\sqrt{ \text{Det}\left( -g^{J}_{\mu\nu} + T_{4}^{-1} \mathcal{F}_{\mu\nu} \right) }~, \label{DBI}
\end{equation}
where $\mathcal{F}_{\mu\nu}$ is the Maxwell tensor 
$\mathcal{F}_{\mu\nu} = \partial_{\mu}A_{\nu} - \partial_{\nu}A_{\mu}$, 
and $\mathcal{T}_{4} = \frac{1}{2\pi\alpha'} = \frac{M_{s}^{2}}{2\pi}$ 
is the (open) string tension, with $\alpha'=M_{s}^{-2}$ the Regge slope and $M_{s}$ the string mass scale, which in general is different from the four-dimensional Planck scale.

Under the assumption of a constant dilaton and flat Minkowski spacetime, the Born–Infeld (BI) parameter $\mathcal{T}_{4}^{2}$ can be constrained in collider experiments through light-by-light scattering, a process that is now experimentally established at the LHC (see \cite{dEnterria:2013zqi,ATLAS:2017fur,CMS:2018erd}). In particular, such scattering studies \cite{Ellis:2017edi} provide a lower bound $\mathcal{T}_{4} \gtrsim 100~\text{GeV}$. Within string theory, this translates into a (weak) lower limit on the string mass scale, $M_{s} \gtrsim 0.25~\text{TeV}$. Current LHC searches in extra-dimensional scenarios further strengthen this bound to $M_{s} \gtrsim \mathcal{O}(10)~\text{TeV}$, with forecasts for future colliders such as the FCC predicting significantly higher limits \cite{Ellis:2022uxv}. Embedding the BI framework into curved spacetime, while fully accounting for dilaton effects, opens an entirely new avenue for testing non-linear electrodynamics through the full machinery of modern gravitational experiments.

Beyond the brane DBI action (\ref{DBI}), one may also consider higher-order derivative corrections in effective low-energy field theories arising from closed strings, such as the heterotic string \cite{Gross:1986mw}. Unlike the DBI brane or open-string case, these corrections do not resum into a closed-form expression for the gauge sector. Nevertheless, extensions of the BI effective action have been proposed in curved $(3+1)$-dimensional spacetimes, incorporating dilaton couplings to the electromagnetic fields in a BI-type non-linear electrodynamics (NED) framework \cite{Yazadjiev:2005za,Dehghani:2006em,Sheykhi:2006dz}. Passing into the Einstein frame in four dimensions, via the transformation of the metric:  $g^{J}_{\mu\nu} \;\to\; g_{\mu\nu} = e^{-2\phi} g^{J}_{\mu\nu}$, we write for the gravitational action (in units 
$c=1$)
\begin{equation}
S = \int d^{4}x \sqrt{-g}
\left[\frac{R}{16\pi G} - \frac{1}{32\pi G}\nabla^{\mu}\Phi \nabla_{\mu}\Phi +\frac{1}{16\pi} \mathcal{L}_{\text{BI}} \right]~,
\end{equation}
\begin{equation}
\mathcal{L}_{\text{BI}} = 4\beta_{\text{BI}} e^{\gamma\Phi}
\left( 1 - \sqrt{ 1 + \frac{e^{-2\gamma\Phi}}{2\beta_{\text{BI}}} \mathcal{F}^{2}
- \frac{e^{-4\gamma\Phi}}{16\beta_{\text{BI}}^{2}} (\mathcal{F}\tilde{\mathcal{F}})^{2} } \right)~,\label{BI}
\end{equation}
where we have redefined the dilaton field $\phi\rightarrow\Phi/2$, $\tilde{\mathcal{F}}_{\mu\nu} = \tfrac{1}{2} \, \epsilon_{\mu\nu\rho\sigma} \, 
\mathcal{F}^{\rho\sigma}$ is the dual of the Maxwell tensor, with $\epsilon_{\mu\nu\rho\sigma}$ the fully antisymmetric Levi-Civita tensor, $\gamma$ denotes the dilaton coupling, and $\beta_{\text{BI}}$ is the generalized BI parameter, with mass dimension $+2$. In the string case, $\beta_{\text{BI}}$ is identified with $\mathcal{T}_{4}^{2}$. To ensure consistency with the $\mathcal{O}(\alpha')$ Maxwell terms of the heterotic string effective action \cite{Gross:1986mw}, namely $e^{-\Phi}\mathcal{F}^{2}$, we will fix $\gamma=1$. 

We seek magnetically charged GM solutions in the following string-inspired gravitational theory:
\begin{equation}
    S = \int d^{4}x \sqrt{-g} \left[\frac{R}{16\pi G} + \mathcal{L}_{\text{GM}} - \frac{1}{32\pi G}\nabla^{\mu}\Phi \nabla_{\mu}\Phi +\frac{1}{16\pi} \mathcal{L}_{\text{BI}} \right]~,\label{model1}
\end{equation}
where $\mathcal{L}_{\text{GM}}$ is the standard GM Lagrangian given in Eq.~\eqref{LGM}. The equations of motion read
\begin{align}
    &\square\chi^{a}=\lambda\chi^{a}\left(\chi^{a}\chi^{a}-\eta^{2}\right)~,\label{HiggsEOM}\\
    &\nabla_{\mu}\left[e^{\Phi}\frac{-e^{-2\Phi}\mathcal{F}^{\mu\nu}+\frac{1}{4\beta_{\text{BI}}^{2}}e^{-4\Phi}(\mathcal{F}_{\rho\sigma}\tilde{\mathcal{F}}^{\rho\sigma})\tilde{\mathcal{F}}^{\mu\nu}}{\sqrt{ 1 + \frac{e^{-2\Phi}}{2\beta_{\text{BI}}} \mathcal{F}^{2} - \frac{e^{-4\Phi}}{16\beta_{\text{BI}}^{2}} (\mathcal{F}\tilde{\mathcal{F}})^{2} }}\right]=0~,\label{BIEOM}\\
    &\square\Phi=-4 G e^{\Phi}\beta_{\text{BI}}\left(1-\sqrt{ 1 + \frac{e^{-2\Phi}}{2\beta_{\text{BI}}} \mathcal{F}^{2} - \frac{e^{-4\Phi}}{16\beta_{\text{BI}}^{2}} (\mathcal{F}\tilde{\mathcal{F}})^{2} }\right)-4 G e^{\Phi}\beta_{\text{BI}}\frac{\frac{e^{-2\Phi}}{2\beta_{\text{BI}}} \mathcal{F}^{2}-\frac{2e^{-4\Phi}}{16\beta_{\text{BI}}^{2}} (\mathcal{F}\tilde{\mathcal{F}})^{2}}{\sqrt{ 1 + \frac{e^{-2\Phi}}{2\beta_{\text{BI}}} \mathcal{F}^{2} - \frac{e^{-4\Phi}}{16\beta_{\text{BI}}^{2}} (\mathcal{F}\tilde{\mathcal{F}})^{2} }}~,\label{DilatonEOM}\\
    &\mathcal{E}_{\mu\nu}:~R_{\mu\nu}-\frac{1}{2}g_{\mu\nu}R=8\pi G T_{\mu\nu}~,\label{EinsteinEOM}
\end{align}
where 
\begin{equation}
    T_{\mu\nu}=T_{\mu\nu}^{\text{Higgs}}+T_{\mu\nu}^{\text{EM}}+T_{\mu\nu}^{\Phi}~,\label{SE}
\end{equation}
with
\begin{equation}
        T_{\mu\nu}^{\text{Higgs}}=g_{\mu\nu}\left(-\frac{1}{2}(\nabla_{\mu}\chi^{a})(\nabla^{\mu}\chi^{a})-\frac{\lambda}{4}\left(\chi^{a}\chi^{a}-\eta^{2}\right)^{2}\right)+\partial_{\mu}\chi^{a}\partial_{\nu}\chi^{a}~,
\end{equation}
\begin{equation}
        T_{\mu\nu}^{\text{EM}}=g_{\mu\nu}\frac{e^{\Phi}}{4\pi}\beta_{\text{BI}}\left( 1 - \sqrt{ 1 + \frac{e^{-2\Phi}}{2\beta_{\text{BI}}} \mathcal{F}^{2} - \frac{e^{-4\Phi}}{16\beta_{\text{BI}}^{2}} (\mathcal{F}\tilde{\mathcal{F}})^{2} } \right)+\frac{e^{\Phi}}{4\pi}\frac{e^{-2\Phi}\mathcal{F}_{\mu}^{~\rho}\mathcal{F}_{\nu\rho}-\frac{e^{-4\Phi}}{2\beta_{\text{BI}}}(\mathcal{F}\tilde{\mathcal{F}})\mathcal{F}_{\nu}^{~\rho}\tilde{\mathcal{F}}_{\mu\rho}}{\sqrt{ 1 + \frac{e^{-2\Phi}}{2\beta_{\text{BI}}} \mathcal{F}^{2} - \frac{e^{-4\Phi}}{16\beta_{\text{BI}}^{2}} (\mathcal{F}\tilde{\mathcal{F}})^{2} }}~,
\end{equation}
\begin{equation}
        T_{\mu\nu}^{\Phi}=-\frac{g_{\mu\nu}}{32\pi G}(\partial\Phi)^{2}+\frac{1}{16\pi G}\partial_{\mu}\Phi\partial_{\nu}\Phi~.
\end{equation}
Seeking for a static and spherically symmetric solution with a magnetic charge, we assume that $\Phi=\Phi(r)$, and introduce the ansätze
\begin{equation}
    ds^{2}=-B(r)dt^{2}+\frac{dr^{2}}{A(r)}+R^{2}(r)\left(d\theta^{2}+\sin^{2}\theta d\phi^{2}\right)~,\label{MetricGlobal}
\end{equation}
\begin{equation}
    \mathcal{F}_{\theta\phi}=-\mathcal{F}_{\phi\theta}=R^{2}(r)W(r)\sin\theta~,\label{EMstrength}
\end{equation}
where all the other components of the EM strength are zero. 

The global O(3) symmetry in Eq.~\eqref{model1} is spontaneously broken due to the Higgs potential. In the broken symmetry phase the Higgs triplet satisfies the equation
\begin{equation}
    \chi^{a}\chi^{a}=\eta^{2}~.
\end{equation}
This condition defines a two-dimensional submanifold of the three-dimensional field space of $\chi^{a}$, which is homeomorphic to $S^{2}$. The three components $\chi^{a}$ lie on a sphere of radius $\eta$, and therefore the vacuum manifold $M_{\text{vac}}\simeq S^2$ has a non-trivial second homotopy group, $\pi_{2}(M_{\text{vac}})=\mathbb{Z}$. Consequently, the vacuum manifold allows for the formation of topologically stable monopole defects. Seeking spherically symmetric solutions, we adopt the following ansatz for the Higgs triplet
\begin{equation}
    \chi^{a}=\eta h(r) \frac{x^{a}}{r}~,~~x^{a}x^{a}=r^{2}~,\label{Higgs}
\end{equation}
such that $h(r) \to 1$ as $r \to \infty$ and $h(r) \to 0$ as $r \to 0$. 

The EM equations of motion are identically satisfied. The remaining equations 
can be solved by specifying the function $W(r)$ in the form 
\begin{equation}
    W(r)=\frac{Q_{m}}{R^{2}(r)}~,
    \label{W}
\end{equation}
which corresponds to the standard magnetic monopole configuration in BI electrodynamics.
To solve the equations of motion, we assume a monopole structure similar to that introduced in Ref.~\cite{Harari:1990cz}, in which the monopole has a core of radius $\delta$. Outside the monopole core ($r>\delta$), the global O(3) symmetry is spontaneously broken, and hence $h(r)=1$. Moreover, far from the core the magnetic field is weak, and Born--Infeld electrodynamics reduces to Maxwell electrodynamics. Therefore, the dilaton equation and the stress-energy tensor of the electromagnetic field approximately read
\begin{equation}
    \square\Phi=- G e^{-\Phi}\mathcal{F}_{\mu\nu}\mathcal{F}^{\mu\nu}~,
\end{equation}
\begin{equation}
        T_{\mu\nu}^{\text{EM}}=-g_{\mu\nu}\frac{e^{-\Phi}}{16\pi}\mathcal{F}_{\mu\nu}\mathcal{F}^{\mu\nu}+\frac{e^{-\Phi}}{4\pi}\mathcal{F}_{\mu}^{~\rho}\mathcal{F}_{\nu\rho}~.
        \label{SEtensorEMlargeR}
\end{equation}
Solving the equations of motion in the region outside the monopole core, we obtain 
\begin{equation}
    B_{\text{ext}}(r)=A_{\text{ext}}(r)=1-8\pi G\eta^{2}-\frac{2GM}{r}~,~~R(r)=\sqrt{r(r-\zeta)}~,\label{metricB}
\end{equation}
\begin{equation}
    \Phi(r)=-\ln\left[1-\frac{\zeta}{r}\right]~,~~Q_{m}=\sqrt{M\zeta}~,\label{dilatonfield}
\end{equation}
where $r\ge \zeta$, which ensures that the geometry remains real. Here $R(r)$ is the areal radius. The quantity 
$M$ is the ADM mass of the monopole, and $\zeta$ is the dilaton scalar charge that controls the $\frac{1}{r}$-term of dilaton in its expansion for large $r$, 
\begin{equation}
\Phi(r)\simeq \frac{\zeta}{r} + \mathcal{O}(r^{-2})\,.
\end{equation}
The magnetic charge of the monopole is denoted by $Q_{m}$, and is determined by both the dilaton scalar charge and the monopole mass, implying that the dilaton field acts as a secondary scalar hair. For a review of hairy black holes, see Ref.~\cite{Herdeiro:2015waa}; for studies of magnetically charged hairy black holes in non-linear electrodynamics, see Refs.~\cite{Karakasis:2022xzm,Theodosopoulos:2023ice}.

The solutions \eqref{metricB} and \eqref{dilatonfield} reduce, for $\eta=0$, to those first obtained in Ref.~\cite{Garfinkle:1990qj}. In Schwarzschild coordinates, where $g_{\theta\theta}=R^{2}$, these solutions read 
\begin{equation}
    B_{\text{ext}}(R)=1-8\pi G\eta^{2}-\frac{4GM}{\zeta+\sqrt{4R^{2}+\zeta^{2}}}~,~~A_{\text{ext}}(R)=\frac{B(R)}{F(R)}~,\label{MetricSchw}
\end{equation}
\begin{equation}
    F(R)=\frac{4R^{2}}{\zeta^{2}+4R^{2}}~,~~\Phi(R)=-\ln\left[1-\frac{2\zeta}{\zeta+\sqrt{4R^{2}+\zeta^{2}}}\right]~.\label{DilatonSchw}
\end{equation}
Inside the monopole core ($r<\delta$), the global O(3) symmetry is unbroken. We assume that the Higgs triplet vanishes in this region, $h(r)=0$. In the spirit of Ref.~\cite{Mavromatos:2016mnj}, we further assume that gravity becomes strong in the core and that the underlying string theory is in a strongly coupled regime. Accordingly, the dilaton is taken to be stabilized at large values near the origin ($\Phi \to +\infty$), such that $\frac{e^{-2\Phi}}{\beta_{\text{BI}}}F_{\mu\nu}F^{\mu\nu}\approx 0$. In this limit, the dilaton equation is trivially satisfied and the electromagnetic sector becomes negligible. Consequently, regular solutions of the Einstein equations can be obtained inside the monopole core. In Schwarzschild coordinates, where $g_{\theta\theta}=R^{2}$, the metric functions are given by
\begin{equation}
    B_{\text{int}}(R)= A_{\text{int}}(R)=1-\frac{2\pi G \lambda\eta^{4}}{3}R^{2}~.\label{coresolution}
\end{equation}
This solution describes a de Sitter monopole core, in agreement with similar constructions in Refs.~\cite{Mavromatos:2018drr,Chatzifotis:2022ubq,Harari:1990cz}. 

To perform the Israel matching at the surface of the de Sitter core of radius $\delta$, and thereby determine both the ADM mass $M$ and the core radius $\delta$, we adopt the approximation introduced in Ref.~\cite{Mavromatos:2018drr}, namely a homogeneous Reissner--Nordstr\"om (RN) metric valid at large $R$:
\begin{equation}
    B_{\text{ext}}(R)\approx A_{\text{ext}}(R)\approx1-8\pi G \eta^{2}-\frac{2GM}{R}+\frac{GM\zeta}{R^{2}}+\mathcal{O}\left(\frac{1}{R^{3}}\right)~.\label{ApprBIsreal}
\end{equation}
The Israel matching conditions at $R=\delta$ read 
\begin{equation}
    B_{\text{ext}}(\delta)=B_{\text{int}}(\delta)~,~~B'_{\text{ext}}(\delta)=B'_{\text{int}}(\delta)~.
\end{equation}
To simplify the analysis, we introduce the dimensionless variables $\tilde{R}=\eta\sqrt{\lambda}R$, $\tilde{\delta}=\eta\sqrt{\lambda}\delta$, $\tilde{\zeta}=\eta\sqrt{\lambda}\zeta$ and $\tilde{M}=\frac{\sqrt{\lambda}}{\eta}M$. Substituting the approximate exterior metric~\eqref{ApprBIsreal} and interior metric~\eqref{coresolution} and using $8\pi G=1$ in reduced Planck units, the two conditions become respectively
\begin{align}
\frac{2\tilde{M}}{\tilde{\delta}}-\frac{\tilde{M}\tilde{\zeta}}{\tilde{\delta}^{2}}
&=
\frac{2\pi\tilde{\delta}^{2}}{3},
\label{eq:Israel1}
\\
\frac{2\tilde{M}}{\tilde{\delta}^{2}}-\frac{2\tilde{M}\tilde{\zeta}}{\tilde{\delta}^{3}}
&=
\frac{4\pi\tilde{\delta}}{3}.
\label{eq:Israel2}
\end{align}
Dividing~\eqref{eq:Israel1} by $\tilde{\delta}$ and comparing with~\eqref{eq:Israel2} fixes $\tilde{M}$ in terms of $\tilde{\delta}$ and $\tilde{\zeta}$. The matching conditions then yield
\begin{equation}
    \tilde{M}=\frac{2\pi}{3}\frac{\tilde{\delta}^{4}}{\tilde{\zeta}-\tilde{\delta}}~,\label{Mass}
\end{equation}
and three real roots for $\tilde{\delta}$, when $\tilde{\zeta}\in\big[0,\frac{3}{2\sqrt{2}}\big]\cup\big[4\sqrt{3},+\infty\big)$:
\begin{align}
    \tilde{\delta}_{1}&=\frac{2}{9}\left(\tilde{\zeta}-\text{sign}[a]\sqrt{\tilde{\zeta}^{2}+27}\left(\cos\left[\frac{1}{3}\tan^{-1}\left[\frac{\sqrt{b}}{|a|}\right]\right]+\sqrt{3}\sin\left[\frac{1}{3}\tan^{-1}\left[\frac{\sqrt{b}}{|a|}\right]\right]\right)\right)~,\\
    \tilde{\delta}_{2}&=\frac{2}{9}\left(\tilde{\zeta}-\text{sign}[a]\sqrt{\tilde{\zeta}^{2}+27}\left(\cos\left[\frac{1}{3}\tan^{-1}\left[\frac{\sqrt{b}}{|a|}\right]\right]-\sqrt{3}\sin\left[\frac{1}{3}\tan^{-1}\left[\frac{\sqrt{b}}{|a|}\right]\right]\right)\right)~,\\
    \tilde{\delta}_{3}&=\frac{2}{9}\left(\tilde{\zeta}+2\text{sign}[a]\sqrt{\tilde{\zeta}^{2}+27}\cos\left[\frac{1}{3}\tan^{-1}\left[\frac{\sqrt{b}}{|a|}\right]\right]\right)~,
\end{align}
where
\begin{equation}
    a=4\tilde{\zeta}^{3}-567\tilde{\zeta}~,~~b=27^{2}(\tilde{\zeta}^{2}-48)(8\tilde{\zeta}^{2}-9)~.\label{ab}
\end{equation}
For $\tilde{\zeta}\in\big[0,\frac{3}{2\sqrt{2}}\big]\cup\big[4\sqrt{3},+\infty\big)$ one finds $b>0$. In the special case $\tilde{\zeta}=0$, the only positive root is $\tilde{\delta}_{1}=2$, yielding a monopole mass $\tilde{M}=-16\pi/3$, which corresponds to the trivial GM solution \cite{Barriola:1989hx}. For $\tilde{\zeta}\in\big(\frac{3}{2\sqrt{2}},4\sqrt{3}\big)$ the solutions are non-physical, as only a single real negative root exists:
\begin{equation}
    \tilde{\delta}=\frac{2}{9}\left(\tilde{\zeta}-2^{2/3}\frac{\tilde{\zeta}^{2}+27}{(-a-\sqrt{-b})^{1/3}}-2^{-2/3}(-a-\sqrt{-b})^{1/3}\right)<0~,
\end{equation}
where $a$ and $b$ are defined in Eq.~(\ref{ab}); in this interval one has $a,b<0$.

For the solution to describe a physically acceptable particle-like configuration, both the de Sitter core radius and the monopole mass must be positive. These conditions are simultaneously satisfied for $\tilde{\zeta}\geq4\sqrt{3}$, where two positive roots for the dimensionless core radius $\tilde{\delta}$ exist. Let us denote them as $\tilde{R}_{dS1}$ and $\tilde{R}_{dS2}$. For $\tilde{\zeta}\geq4\sqrt{3}$, they are given by
\begin{equation}
    \tilde{R}_{\text{dS1}}=
    \begin{cases} 
        \tilde{\delta}_{1} & \text{if } ~4\sqrt{3}\leq\tilde{\zeta}\leq\frac{9\sqrt{7}}{2} \\
        \tilde{\delta}_{3} & \text{if }~ \frac{9\sqrt{7}}{2}<\tilde{\zeta} 
    \end{cases}~,\label{sector1}
\end{equation}
\begin{equation}
    \tilde{R}_{\text{dS2}}=\tilde{\delta}_{2}~,~~\tilde{\zeta}>4\sqrt{3}~.\label{sector2}
\end{equation}
These solutions satisfy $\tilde{R}_{\text{dS2}}\in[\sqrt{6},2\sqrt{3}]$, $\tilde{R}_{\text{dS1}}\in[2\sqrt{3},+\infty)$, and $\tilde{M}\in[0,+\infty)$. 

\begin{figure}
    \centering
    \includegraphics[width=0.32\textwidth]{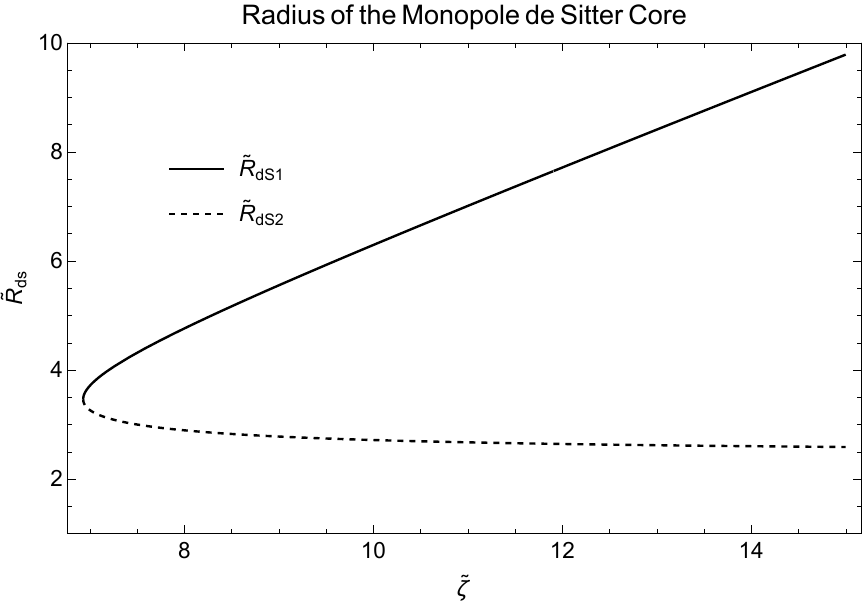}
    \includegraphics[width=0.33\textwidth]{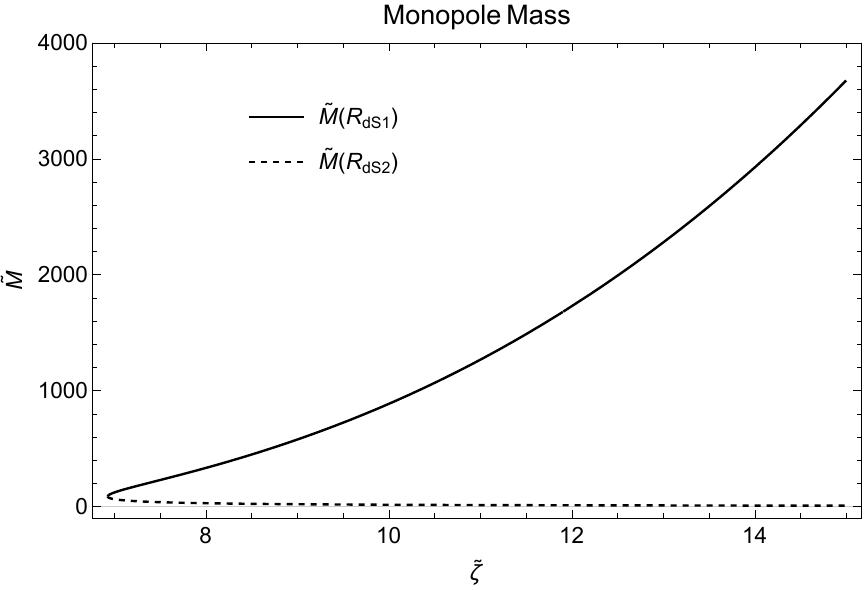}
    \includegraphics[width=0.33\textwidth]{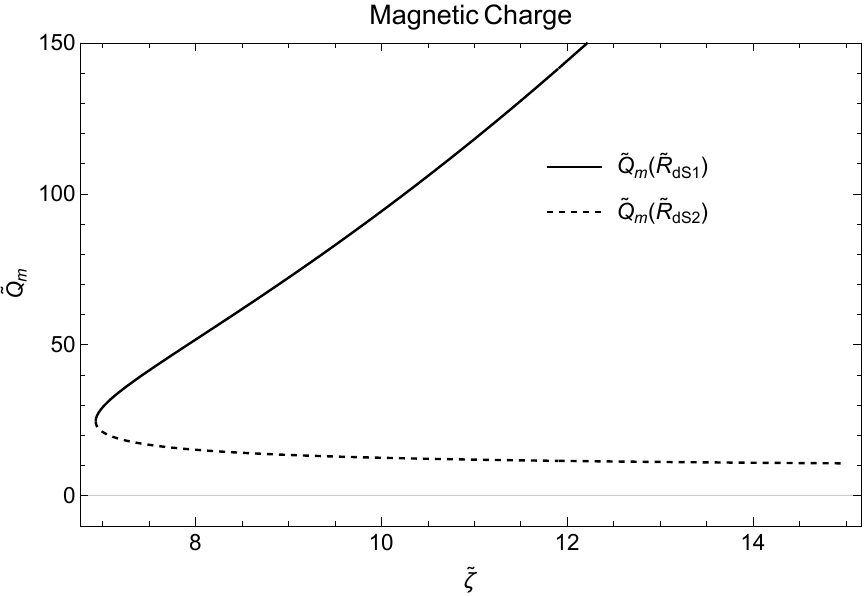}
    \caption{Dimensionless de Sitter core radius $\tilde{R}_{\text{dS}}$ (left), monopole mass $\tilde{M}$ (middle), and magnetic charge $\tilde{Q}_{m}$ (right) as functions of the dimensionless dilaton charge $\tilde{\zeta}$. Solid lines correspond to the solution branch $\tilde{R}_{\text{dS1}}$ [Eq.~(\ref{sector1})], while dashed lines correspond to $\tilde{R}_{\text{dS2}}$ [Eq.~(\ref{sector2})].}
    \label{fig:1}
\end{figure}

\begin{figure}
    \centering
    \includegraphics[width=0.45\textwidth]{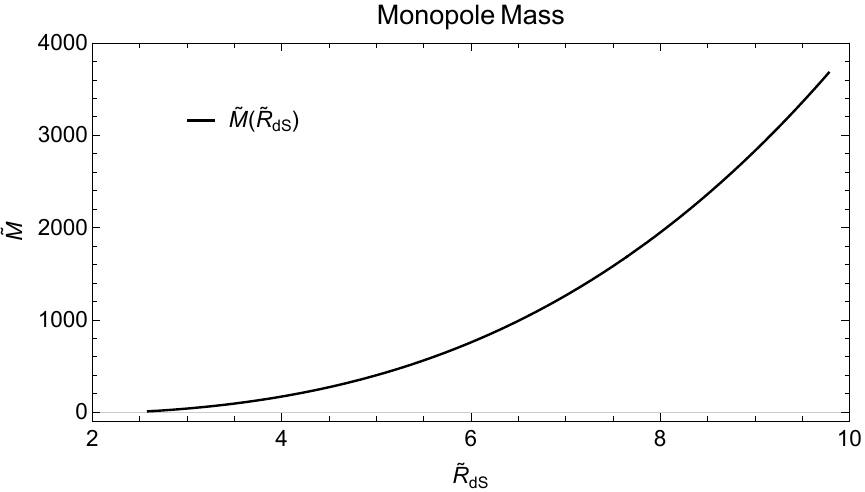}
    \includegraphics[width=0.45\textwidth]{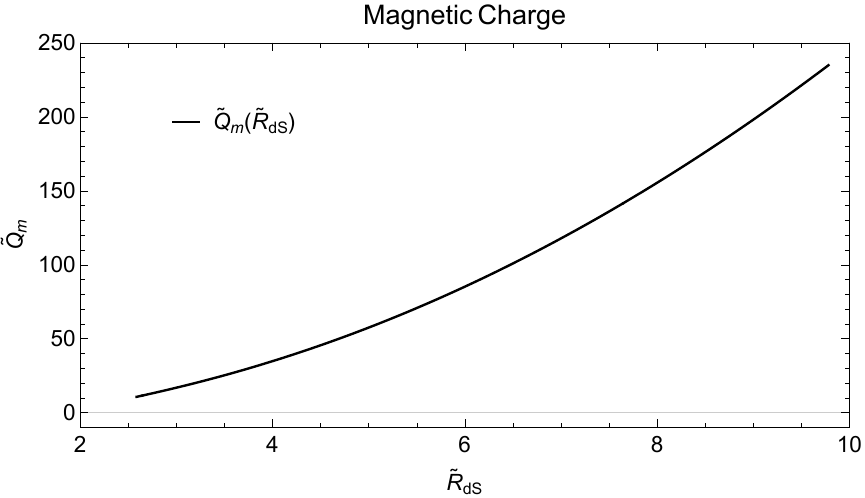}
    \caption{Dimensionless monopole mass $\tilde{M}$ (left) and magnetic charge $\tilde{Q}_{m}$ (right) as functions of the de Sitter core radius $R_{\text{dS}}$.}
    \label{fig:2}
\end{figure}

In Fig.~\ref{fig:1}, we plot the de Sitter core radius $R_{\text{dS}}$, the monopole mass $\tilde{M}$, and the magnetic charge $\tilde{Q}_{m}=\sqrt{\tilde{\zeta}\tilde{M}}$ as functions of the dilaton charge $\tilde{\zeta}$.
In Fig.~\ref{fig:2}, the monopole mass and magnetic charge are shown as functions of the de Sitter core radius $R_{\text{dS}}$. Both $\tilde{M}(R_{\text{dS}})$ and $\tilde{Q}_{m}(R_{\text{dS}})$ are monotonically increasing functions of $\tilde{R}_{\text{dS}}$.

\begin{quote}
\textbf{Remark (minimum magnetic charge).} In the limiting case of the smallest monopole, $\tilde{R}_{\mathrm{dS}}\to\sqrt{6}$ (the lower end of the $\tilde{R}_{\mathrm{dS2}}$ branch), the dimensionless mass vanishes $\tilde{M}\to 0$ while the magnetic charge $\tilde{Q}_m=\sqrt{\tilde{\zeta}\tilde{M}}$ approaches a finite nonzero value. This implies the existence of a \emph{minimum nonzero magnetic charge} in this model: a zero-mass limit exists in which the monopole is held up entirely by its magnetic charge and the de~Sitter core pressure, with no gravitational contribution.
\end{quote}

\begin{figure}
    \centering
    \includegraphics[width=0.45\textwidth]{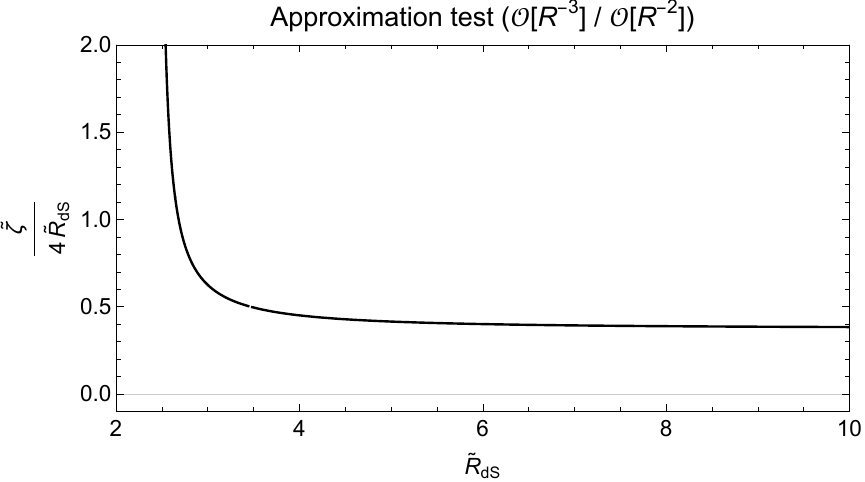}
    \includegraphics[width=0.45\textwidth]{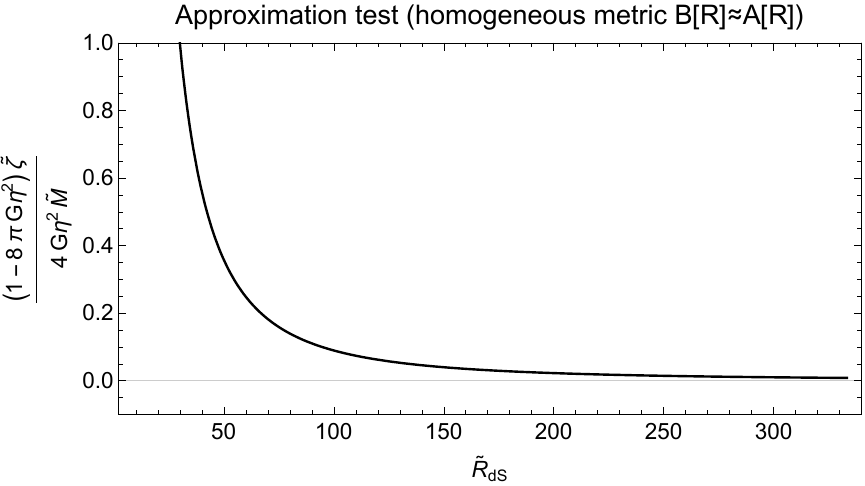}
    \caption{Left: Ratio of the $R^{-3}$ to $R^{-2}$ terms in the large-$R$ expansion of the metric function $B(R)$ in Schwarzschild coordinates [Eq.~(\ref{ApprxB})], plotted as a function of the de Sitter core radius $R_{\text{dS}}$. Right: Relative difference between the Reissner--Nordstr\"om contributions to the metric functions $B(R)$ and $A(R)$, defined as $(A(R)-B(R))/B(R)$, as a function of $R_{\text{dS}}$.}
    \label{fig:3}
\end{figure}

Having determined the monopole mass and the radius of the de Sitter core, we now assess the validity of the approximations used in the Israel matching. Outside the monopole core, in the large-$R$ region, the metric components read
\begin{align}
    B_{\text{ext}}(R)&\approx1-8\pi G \eta^{2}-\frac{2GM}{R}+\frac{GM\zeta}{R^{2}}-\frac{GM\zeta^{2}}{4R^{3}}+\mathcal{O}\left(\frac{1}{R^{5}}\right)~,\label{ApprxB}\\
    A_{\text{ext}}(R)&\approx1-8\pi G \eta^{2}-\frac{2GM}{R}+\frac{GM\zeta+(1-8\pi G \eta^{2})\frac{\zeta^{2}}{4}}{R^{2}}+\mathcal{O}\left(\frac{1}{R^{3}}\right)~.\label{ApprxA}
\end{align}
In deriving the analytic expressions for the monopole mass and core radius, we retained terms up to the Reissner--Nordstr\"om (RN) contribution in the metric function $B(R)$. To assess the validity of this approximation, we plot in the left panel of Fig.~\ref{fig:3} the ratio of the $R^{-3}$ and $R^{-2}$ terms in Eq.~(\ref{ApprxB}), evaluated at $R=R_{\text{dS}}$. For small monopoles, this ratio becomes significantly larger than unity, indicating that higher-order terms cannot be neglected. For large monopoles, the ratio approaches a lower bound of $3/8$. Overall, the RN approximation is unreliable for small monopoles, where the mass and charge may deviate substantially from the analytic estimates; for larger monopoles, the approximation improves, although it remains approximate and may still lead to quantitative discrepancies.

In addition, we assumed a homogeneous metric outside the monopole core. For this approximation to hold up to the RN order, the metric functions $B(R)$ and $A(R)$ must coincide up to terms of order $R^{-2}$, which requires $G\tilde{M}\gg(1-8 \pi G \eta^{2})\frac{\tilde{\zeta}}{4\eta^{2}}$. In the right panel of Fig.~\ref{fig:3}, we plot the ratio $(1-8 \pi G \eta^{2})\frac{\tilde{\zeta}}{4\eta^{2}G\tilde{M}}$ as a function of $R_{\text{dS}}$, in units $G=1$ and for $\eta=10^{-2}$. This ratio decreases to zero as $R_{\text{dS}}\to\infty$, demonstrating that the homogeneous metric approximation becomes increasingly accurate for large monopoles. 

Although the analytic approximations break down in part of the parameter space, in particular in the small-mass regime, we now verify numerically that magnetic monopole solutions still exist and determine their masses in cases where the analytic treatment becomes unreliable. As a representative example, we consider  
\begin{equation}\label{betalambda}
    \eta=10^{-2}~,~~\lambda=10^{4}~,~~ \beta_{BI}=1~,~~\zeta=\frac{4\sqrt{3}}{\eta\sqrt{\lambda}}~,
\end{equation}
and we work in units $G=c=1$. 
The BI parameter is related to the string scale $M_{s}$ via $\beta_{\text{BI}}\approx M_{s}^{4}$. Identifying $M_s$ with the (reduced) Planck scale, 
\begin{align}\label{MsMPl}
M_s \simeq M_{\rm Pl}\,,
\end{align}
the BI parameter reads 
\begin{align}
\label{b1}
\beta_{\text{BI}}=1\,,
\end{align}
in (reduced) Planck units, where:
\begin{align}\label{Gunits}
    8\pi\, G = 1\,.
\end{align}

A numerical analysis of the standard GM \cite{Barriola:1989hx} has been performed in Ref.~\cite{Harari:1990cz}, using the Runge-Kutta method. The authors of Ref.~\cite{Chatzifotis:2022ubq} also find a numerical solution of a GM associated with the spontaneous breaking of a global O(3) symmetry within an extended Gauss Bonnet theory of gravity, which includes a dilaton field. The presence of the dilaton and Higgs fields makes the numerical problem stiff, and thus certain numerical methods for solving the equations of motion can become numerically unstable unless an extremely small step size is used. We remind the reader that an ordinary differential equation problem is considered stiff if the desired solution changes slowly while nearby solutions change rapidly, necessitating small steps for the numerical method to achieve satisfactory results. This is also the case with the dilaton and Higgs fields in our work. Therefore, we will solve the differential problems exploiting the ``StiffnessSwitching" method of Mathematica, which uses a pair of extrapolation methods as the default. Specifically, the stiff solver uses the Linearly Implicit Euler method, while the non-stiff solver uses the Explicit Modified Midpoint method.

The equations of motion are given by Eqs.~(\ref{HiggsEOM})--(\ref{EinsteinEOM}). The ansatz (\ref{EMstrength}) automatically satisfies the BI equations of motion. Additionally, the remaining five independent equations of motion yield $W(r)=Q_{m}/R^{2}(r)$.
To ensure that the equations of motion are satisfied in the large-$r$ region, the magnetic charge must satisfy $Q_{m}=\sqrt{M\zeta}$, where $M$ is the monopole mass and $\zeta$ is the dilaton charge. We work in the coordinate system $(t,r,\theta,\phi)$, with the metric given by Eq.~(\ref{MetricGlobal}) and $R(r)=\sqrt{r(r-\zeta)}$ for $r\ge\zeta$. 

To obtain the numerical solutions, we solve the remaining four independent equations of motion for $B'(r)$, $A'(r)$, $h''(r)$, and $\Phi''(r)$, where the prime denotes differentiation with respect to the radial coordinate $r$. These equations are solved simultaneously. We do not display their explicit form, as the resulting expressions are lengthy and not particularly illuminating.  

To avoid singular behavior, we impose initial conditions at $r=\zeta+\epsilon$ with $\epsilon=10^{-6}$:
\begin{equation}
    \begin{split}
        &h'(\zeta+\epsilon)\approx4\cdot 10^{4}~,~~h(\zeta+\epsilon)\approx h'(\zeta+\epsilon)\epsilon~,~~\Phi'(\zeta+\epsilon)\approx-10^{6}~,\\
        &\Phi(\zeta+\epsilon)\approx13~,~~B(\zeta+\epsilon)\approx 1~,~~A(\zeta+\epsilon)\approx 10^{-2}~.
    \end{split}
\end{equation}
These initial conditions are chosen to reproduce the correct asymptotic behavior. 
They are also consistent with the expected structure inside the dS core: for the chosen coordinate system, the interior metric functions read $B_{\text{int}}(r)=1-\frac{2\pi G \lambda\eta^{4}}{3}r(r-\zeta)$ and $A_{\text{int}}(r)=B_{\text{int}}(r)/R'^{2}(r)$, implying $B(\zeta+\epsilon)\approx1$ and $A(\zeta+\epsilon)\approx 0$ near the origin. In Schwarzschild coordinates, both metric functions remain regular, ensuring the absence of curvature singularities. 
The Higgs field is approximately zero near the origin, where the global O(3) symmetry is unbroken, and rapidly approaches unity outside the monopole core, where the symmetry is broken. Accordingly, its radial derivative is large near the origin. The initial conditions for the dilaton are chosen to reproduce the asymptotic profile $\Phi(r) = -\ln\left(1 - \frac{\zeta}{r}\right)$ at large $r$. 
In addition, we impose that the mass function
\begin{equation}
    M(r)=\frac{r}{2G}\left(1-8\pi G\eta^{2}-B(r)\right)~,\label{massfunction}
\end{equation}
approaches a constant as $r \to \infty$. The asymptotic value of $M(r)$ then corresponds to the ADM mass of the monopole. 

\begin{figure}
    \centering
    \includegraphics[width=0.45\textwidth]{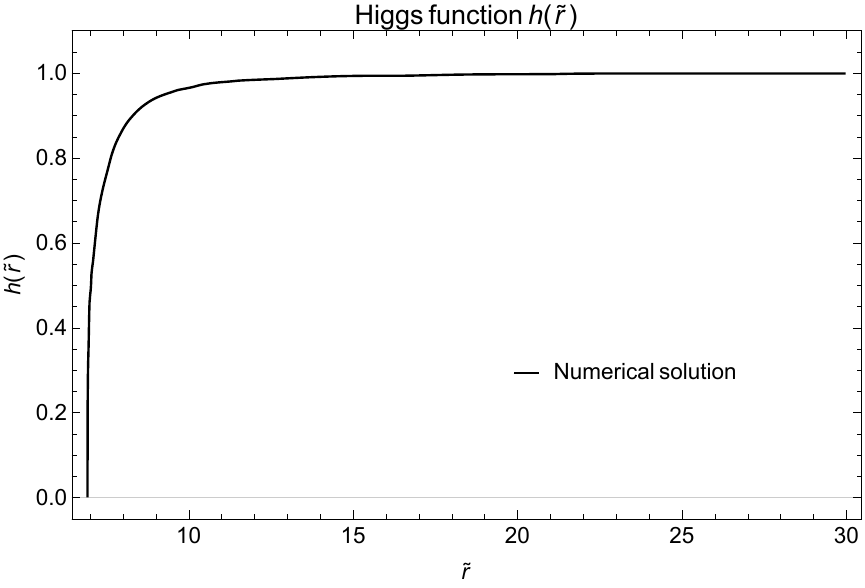}
    \includegraphics[width=0.45\textwidth]{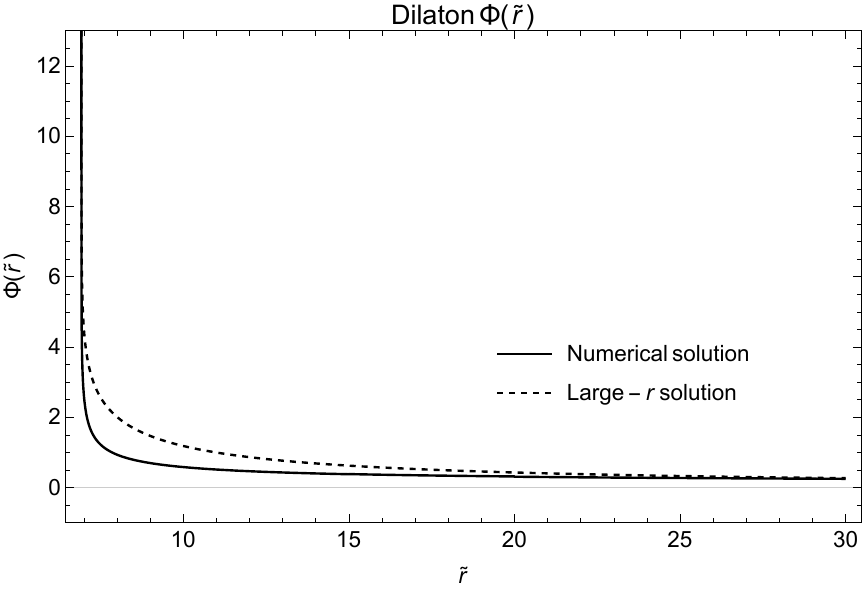}
    \includegraphics[width=0.46\textwidth]{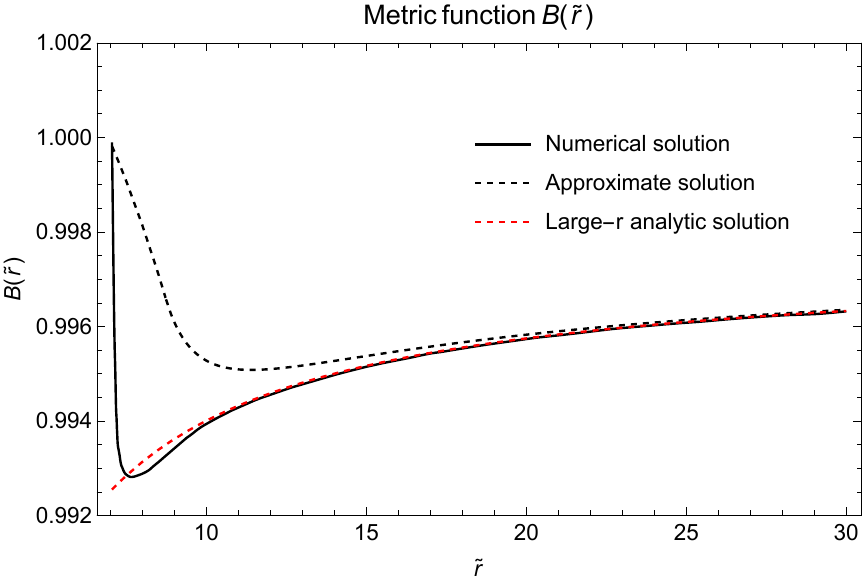}
    \includegraphics[width=0.45\textwidth]{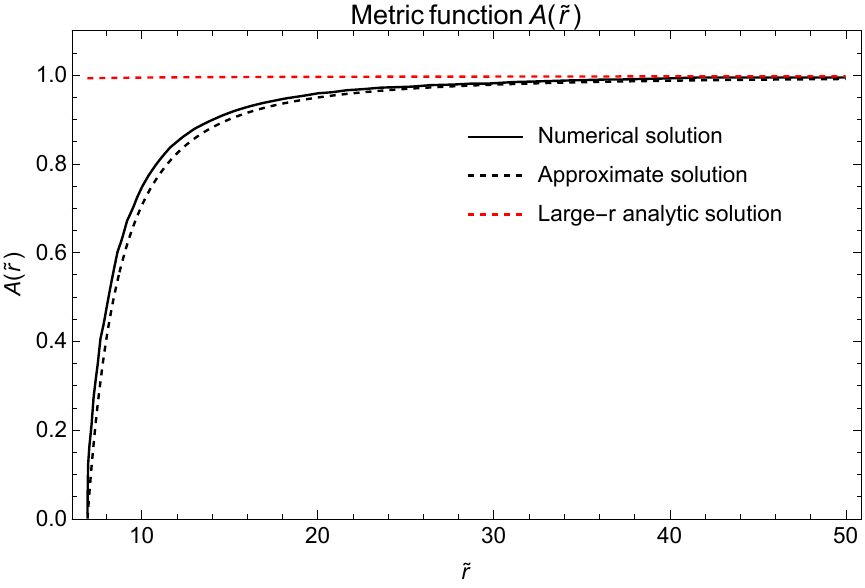}
    \caption{Numerical solutions of the equations of motion. Top-left: Higgs function. Top-right: Dilaton field (numerical solution shown by the solid line, analytic large-$r$ solution by the dashed line). Bottom panels: Metric functions $B(\tilde{r})$ (left) and $A(\tilde{r})$ (right), including the numerical solutions (solid black), the Israel-matching approximations (dashed black), and the analytic large-$r$ solutions (dashed red).}
    \label{fig:4}
\end{figure}

In Fig.~\ref{fig:4}, we present the numerical solutions and compare them with the corresponding analytic results. 
For the chosen parameter set, the Israel matching predicts a de Sitter core radius $r_{\text{dS}}\approx\frac{9}{\eta\sqrt{\lambda}}$. 
The Higgs profile (top-left panel) increases rapidly in the vicinity of $r_{\text{dS}}$ and approaches unity for $r>r_{\text{dS}}$, indicating that the global O(3) symmetry is unbroken inside the core and broken outside. 
This behavior is consistent with previous studies of global monopoles~\cite{Harari:1990cz,Chatzifotis:2022ubq}. 
The dilaton field (top-right panel) attains large values near the origin and reproduces the analytic large-$r$ profile given in Eq.~(\ref{dilatonfield}). Deviations between the numerical and analytic solutions appear in the transition region near the monopole shell, where the approximations are not valid. 
The metric function $B(r)$ (bottom-left panel) remains regular at the origin and approaches the analytic solution~(\ref{metricB}) outside the monopole core. The approximation based on Israel matching is also shown for comparison. 
The metric function $A(r)$ (bottom-right panel) exhibits the expected asymptotic behaviour, approaching the analytic large-$r$ solution $A_{\text{ext}}(r)=B_{\text{ext}}(r)$. For comparison, we also display the approximate solution
$A_{\text{appr}}(r)=\frac{4r(r-\zeta)}{(2r-\zeta)^{2}}\,A(R(r))$,
where $A(R)$ is obtained from the Israel matching. In the bottom panels of Fig.~\ref{fig:4}, the approximate solutions for the metric functions closely follow the numerical profiles.

\begin{figure}
    \centering
    \includegraphics[width=0.5\textwidth]{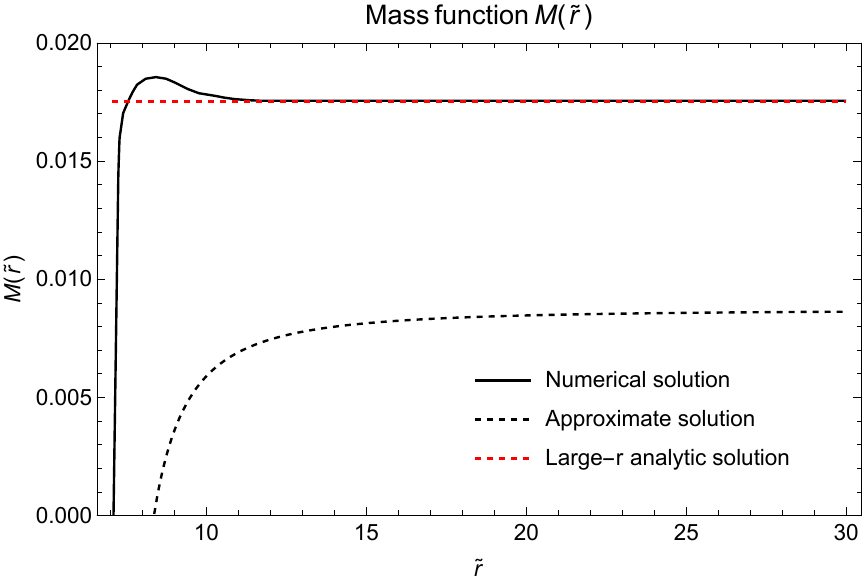}
    \caption{Mass function $M(\tilde{r})$ as a function of the dimensionless radius $\tilde{r}$. The numerical solution (solid black), the Israel-matching approximation (dashed black), and the analytic large-$r$ solution (dashed red) are shown. The mass function approaches a constant value---the ADM mass of the monopole---at large $r$. The numerical and approximate predictions for the monopole mass differ by approximately a factor of two.}
    \label{fig:5}
\end{figure}

In Fig.~\ref{fig:5}, we show the numerical mass function defined in Eq.~(\ref{massfunction}). 
Outside the monopole core, $M(r)$ follows the analytic behavior implied by the metric function $B(r)$ in Eq.~(\ref{metricB}). 
The corresponding Reissner--Nordstr\"om approximation used in the Israel matching is also displayed for comparison. 
The numerical solution approaches a constant as $r\to\infty$, which corresponds to the ADM mass of the monopole. 
For $\tilde{\zeta}=4\sqrt{3}$, the Israel matching predicts $Q_{m}^{\text{Is}}\approx\frac{24}{\sqrt{\lambda}}$ and $M^{\text{Is}}\approx87\frac{\eta}{\sqrt{\lambda}}$. 
The numerical solution yields $Q_{m}\approx\sqrt{2}\;Q_{m}^{\text{Is}}$ and $M\approx2\;M^{\text{Is}}$. 
Thus, the numerical results confirm the existence of magnetic monopole solutions and show that the analytic estimates capture the correct order of magnitude for both the mass and the magnetic charge.

Finally, whereas the self-energy of a monopole diverges in standard Maxwell electrodynamics, it remains finite in Born--Infeld electrodynamics. To illustrate this, we work in Schwarzschild coordinates with $R(r)=r$ for $r\geq 0$. In Maxwell theory, the Lagrangian density behaves as $-\mathcal{F}_{\mu\nu}\mathcal{F^{\mu\nu}}=-\frac{2Q_m^{2}}{r^{4}}$, which leads to a divergent energy density near the origin and hence an infinite self-energy. 
By contrast, in the BI case, the Lagrangian density near the origin scales as $\mathcal{L}_{\text{BI}}\approx-\frac{\sqrt{\beta_{\text{BI}}}}{4\sqrt{2}\pi}\sqrt{\mathcal{F}_{\mu\nu}\mathcal{F^{\mu\nu}}}=-\frac{\sqrt{\beta_{\text{BI}}}}{4\pi}\frac{Q_m}{r^{2}}$. The self-energy in this region is obtained by integrating the Hamiltonian density over the small-radius volume, behaves as $E_{\text{BI}}(r\ll1)=Q_m\sqrt{\beta_{\text{BI}}}\int dr$, and is therefore finite. 
Furthermore, integrating the Hamiltonian density over the entire space, using the parameter set adopted in the numerical analysis and the numerical dilaton profile $\Phi(r)$, yields a total monopole self-energy $E_{\text{BI}}\approx 0.7$, confirming the finiteness of the solution. 

To summarise, the global monopole model with dilaton and Born–Infeld interactions admits self-gravitating magnetic monopole solutions with positive ADM mass, where the magnetic charge is not an independent parameter but is tied to the dilaton charge and the monopole mass. The approximate analytic construction, based on a de Sitter core matched to an exterior magnetically charged geometry, captures the main qualitative features of the solution and identifies the parameter regime in which physically acceptable monopoles exist. Although the Israel matching and Reissner--Nordstr\"om-type approximation are not uniformly accurate across the full parameter space, the numerical analysis confirms the existence of the solutions and shows that the predicted masses and magnetic charges are correct at the level of order of magnitude. An additional appealing feature of the model is that the Born–Infeld sector renders the monopole self-energy finite, in contrast to the Maxwell case. Altogether, these results support the picture that string-inspired dilaton and Born–Infeld couplings provide a natural framework in which global monopoles can be promoted to physically viable magnetic monopoles.

\section{Dyons from Global Monopoles with Dilaton and Axion Charges}\label{sec:GlobalAxion}

In this section we extend the previous construction by incorporating a Kalb–Ramond (KR) field, whose four-dimensional dual description introduces an axion degree of freedom. This additional sector naturally arises in string-inspired frameworks and leads to a non-trivial coupling between the axion and the electromagnetic field through the topological term $\mathcal{F}\tilde{\mathcal{F}}$. As a result, the system admits configurations carrying both electric and magnetic charges, i.e. dyonic solutions. 
Our aim is to investigate whether global monopoles coupled to gravity, a dilaton, and a KR axion field can give rise to self-gravitating dyon solutions with well-defined ADM mass and finite self-energy. We first formulate the effective action and corresponding equations of motion, and then construct static, spherically symmetric solutions by analyzing separately the large- and small-radius regimes, following the approach developed in the previous section.

In addition to the model of the previous section, we introduce a Kalb-Ramond (KR) field with dynamics described by the following Lagrangian
\begin{equation}
    \mathcal{L}^{\text{KR}}=\frac{\sqrt{-g}}{16\pi G}\bigg(\frac{e^{-2\Phi}}{6}H^{\rho\mu\nu}H_{\rho\mu\nu}\bigg)~,
    \label{LagrangianKR}
\end{equation}
where $H_{\rho\mu\nu}$ is the KR field strength tensor, which is associated with the KR field $\mathcal B_{\mu\nu}$ via $H_{\rho\mu\nu}=\partial_{\rho} \mathcal B_{\mu\nu}+\partial_{\mu} \mathcal B_{\nu\rho}+\partial_{\nu} \mathcal B_{\rho\mu}$. The KR field satisfies the following equations of motion
\begin{equation}
    \nabla_{\rho}\big(e^{-2\Phi}H^{\rho\mu\nu}\big)=0~,
    \label{KRequation}
\end{equation}
and the Bianchi identity $\varepsilon^{\mu\nu\rho\sigma}\nabla_{\sigma}H_{\mu\nu\rho}=0$. In 4-dimensions the KR field strength is dual to an axion field $b$ 
\begin{equation}
    H_{\mu\nu\rho}=e^{2\Phi}\varepsilon_{\mu\nu\rho}^{~~~~\sigma}\partial_{\sigma}b~.
    \label{KRdual}
\end{equation}
In superstring theory, anomaly cancellation requires a modification of the KR field strength by appropriate gauge and Lorentz Chern-Simons terms. Thus, as discussed in \cite{Svrcek:2006yi}, the Bianchi identity is modified as
\begin{equation}
    \varepsilon^{\mu\nu\rho\sigma}\nabla_{\sigma}H_{\mu\nu\rho}=\frac{1}{16\pi}(R_{\mu\nu\rho\sigma}\tilde{R}^{\mu\nu\rho\sigma}-\mathcal{F}_{\mu\nu}\tilde{\mathcal{F}}^{\mu\nu})~.
    \label{KRBianchi}
\end{equation}
The KR field strength given in Eq. (\ref{KRdual}) identically satisfies the KR equations of motion (\ref{KRequation}). The complete action incorporating the dynamics of the axion field reads
\begin{equation}
    S = \int d^{4}x \sqrt{-g} \left[\frac{R}{16\pi G} + \mathcal{L}_{\text{GM}} - \frac{1}{32\pi G}\nabla^{\mu}\Phi \nabla_{\mu}\Phi +\frac{1}{16\pi} \mathcal{L}_{\text{BI}}-\frac{e^{2\Phi}}{32\pi G}\nabla_{\mu}b\nabla^{\mu}b+\frac{b}{16\pi G}(R_{\mu\nu\rho\sigma}\tilde{R}^{\mu\nu\rho\sigma}-\mathcal{F}_{\mu\nu}\tilde{\mathcal{F}}^{\mu\nu}) \right]~.\label{model2}
\end{equation}


The equation of motion of the KR axion field associated with this action coincides with the Bianchi identity given by Eq. (\ref{KRBianchi}) and reads
\begin{equation}
    \square b= R_{\mu\nu\rho\sigma}\tilde{R}^{\mu\nu\rho\sigma}-\mathcal{F}_{\mu\nu}\tilde{\mathcal{F}}^{\mu\nu}~.
\end{equation}
To construct a static, spherically symmetric solution with both electric and magnetic charges, we consider that the axion depends only on the radial coordinate $b=b(r)$. For static and spherically symmetric spacetimes one has 
$R_{\mu\nu\rho\sigma}\tilde{R}^{\mu\nu\rho\sigma}=0$, 
while in the presence of both electric and magnetic fields 
$\mathcal{F}_{\mu\nu}\tilde{\mathcal{F}}^{\mu\nu}\neq 0$. 
We further impose a monopole configuration analogous to that of the previous section, and therefore divide the analysis into large- and small-$r$ regimes. In the large-$r$ region, where the global O(3) symmetry is broken, we assume that the BI electrodynamics coincide with the standard Maxwell electrodynamics, and the EM field equations read
\begin{equation}
    \nabla_{\mu}(e^{-\Phi}\mathcal{F}^{\mu\nu}+b\tilde{\mathcal{F}}^{\mu\nu})=0~,
\end{equation}
which is identically satisfied, as shown in \cite{Sur:2005pm}, by considering the following non-vanishing components of the EM field strength
\begin{equation}
    \mathcal{F}_{\theta\phi}=-\mathcal{F}_{\phi\theta}=Q_{m}\sin\theta~~\text{and}~~~\mathcal{F}_{tr}=-\mathcal{F}_{rt}=e^{\Phi(r)}\frac{Q_{e}-(b(r)-b_{0})Q_{m}}{R^{2}(r)},
    \label{EMstrengthAxion}
\end{equation}
where $Q_{e}$ and $Q_{m}$ are the electric and magnetic charges of the dyon respectively, and $b_{0}$ is the asymptotic value for the axion field at spatial infinity. The
combination
\begin{equation}
\Qeff(r):=Q_e-(b(r)-b_0)Q_m
\label{eq:Qeff}\,,
\end{equation}
appearing in \eqref{EMstrengthAxion},
defines an effective electric charge: it vanishes in the purely magnetic limit,
either when $Q_e=0$ and $b(r)=b_0$, or more generally when the axion dressing exactly
compensates the bare electric charge.

For the metric ansatz in Eq. (\ref{MetricGlobal}) the equations of motion are solved by  
\begin{equation}
    B_{\text{ext}}(r)=A_{\text{ext}}(r)=1-8\pi G\eta^{2}-\frac{2GM}{r}~,~~R(r)=\sqrt{r(r-\zeta)}~,
\end{equation}
\begin{equation}
    \Phi(r)=-\ln\left[\frac{Q^{2}R^{2}(r)}{Q^{2}r^{2}-Q_{e}^{2}\zeta(2r-\zeta)}\right]~,~~b(r)=b_{0}+\frac{\zeta Q_{e}Q_{m}(2r-\zeta)}{(r-\zeta)^{2}Q_{e}^{2}+r^{2}Q_{m}^{2}}~,\label{dil2}
\end{equation}
where $r>\zeta$, $M$ is the monopole mass, and 
\begin{equation}
    Q^{2}=Q_{e}^{2}+Q_{m}^{2}=M\zeta~.
\end{equation}
The above solution generalizes the magnetic monopole case of the previous section, with the replacement $Q_m \to Q=\sqrt{Q_e^2+Q_m^2}$. 

In the small-$r$ region, the interior solution coincides with that of the previous section, as long as the dilaton is stabilized at a very large value and the axion is zero inside the monopole core. Thus, we fix $b_{0}=-\frac{Q_{e}}{Q_{m}}$ in order to obtain $b(\zeta)=0$, and the solution inside the monopole core is given by Eq. (\ref{coresolution}). The process of the Israel matching at the monopole core and the numerical calculations are similar to those performed in the previous section, as long as we replace $Q_{m}$ with $Q$, and hence they are omitted. 

In summary, the inclusion of the Kalb–Ramond axion field leads to dyonic generalizations of the global monopole solutions studied previously. The axion coupling induces an effective electric charge through its interaction with the electromagnetic sector, resulting in a non-trivial dressing of the electric component of the field strength. The resulting solutions are characterized by a combined charge $Q^{2}=Q_{e}^{2}+Q_{m}^{2}$, which enters the geometry in a manner analogous to the purely magnetic case, while the axion field interpolates between distinct asymptotic values and vanishes inside the monopole core. The interior solution remains unchanged under appropriate boundary conditions, and the matching procedure is analogous to the monopole case. Overall, these results demonstrate that the framework naturally accommodates self-gravitating dyon solutions, with the axion field playing a central role in relating the electric charge to the underlying topological and scalar structure of the system.

\section{Stability Analysis of the Solutions}\label{sec:stab}

 We next discuss the stability of the MM and Dyon solutions. We examine two types of stability. Mechanical one, and dynamical stability, under linearized perturbations of the solution (linear stability).
We commence the analysis by examining  mechanical stability criteria and the satisfaction of energy conditions of the self-gravitating solutions. 
 
\subsection{Mechanical Stability Criteria and Energy Theorems}
\label{sec:mechstab}

In this section, we  analyze the mechanical stability of the magnetic monopole and dyon solutions obtained in the previous sections and examine the associated energy conditions.  These provide important diagnostics of the physical viability of the solutions, exploring the behaviour of the stress--energy tensor and ensuring the absence of pathologies such as negative energy densities or instabilities.

To define and perform the mechanical stability tests and examine the energy conditions, we write the stress-energy tensor in the physical coordinate system $(t,R,\theta,\varphi)$ in the form of an anisotropic fluid:
\begin{equation}
T^{\mu\nu} = (\rho_E+p_\theta)u^\mu u^\nu + (p_R-p_\theta)n^\mu n^\nu + p_\theta g^{\mu\nu}~,\label{SEtensor}
\end{equation}
where $\rho_E$ denotes the energy density of the fluid measured by a comoving observer, $p_R$ the radial pressure, and $p_\theta$ the tangential pressure. 
The vector $u^\mu$ is the timelike four-velocity of the fluid, while $n^\mu$ is a spacelike unit vector orthogonal to $u^\mu$ and to the angular directions. These vectors satisfy
\begin{equation}
u^\mu = u(R)\,\delta^\mu{}_0, \qquad u^\mu u^\nu g_{\mu\nu}=-1,\label{u}
\end{equation}
\begin{equation}
n^\mu = n(R)\,\delta^\mu{}_1, \qquad n^\mu n^\nu g_{\mu\nu}=1.\label{n}
\end{equation}
Using Eqs.~(\ref{SE})--(\ref{MetricGlobal}) and (\ref{SEtensor})--(\ref{n}), one finds
\begin{equation}
\rho_E = -T^t{}_t~,~~p_R = T^R{}_R~,~~p_\theta = T^\theta{}_\theta~.
\end{equation} 
For an anisotropic fluid of the form \eqref{SEtensor}, the energy conditions are given by
\begin{itemize}
\item Null Energy Conditions (NEC): \qquad $\rho_E+p_R \ge 0 \quad \& \quad \rho_E+p_\theta \ge 0,$
\item Weak Energy Conditions (WEC): \qquad NEC $\quad \& \quad \rho_E \ge 0,$
\item Strong Energy Conditions (SEC): \qquad NEC $\quad \& \quad \rho_E+p_R+2p_\theta \ge 0.$
\end{itemize}

Regarding the mechanical stability criteria, following Ref.~\cite{Farakos:2025byy}, we examine some generalized local Laue stability conditions. 
The Laue condition \cite{Laue:1911lrk} stands that, in the case of an object with a core, the exterior and interior forces should be in balance, or in other words the spatial integral of the fluid pressure $p(R)$ in the entire space should be zero
\begin{equation}
\int_0^\infty dR\, R^2 p(R) = 0~.
\end{equation}
To introduce a stronger mechanical stability criterion, the authors of Ref.~\cite{Farakos:2025byy} discriminate the stress-energy tensor in two parts, one corresponding to long range (LR) force components (EM field) and one corresponding to short range (SR) contributions (the rest of the fields). Thus, the LR and SR pressures can be defined as follows
\begin{equation}
p^{\text{LR}}_R = T^{\text{EM}}{}^R{}_R~,~~p^{\text{LR}}_\theta = T^{\text{EM}}{}^\theta{}_\theta
\end{equation}
and 
\begin{equation}
p^{\text{SR}}_R = T^R{}_R-T^{\text{EM}}{}^R{}_R~,~~p^{\text{SR}}_\theta = T^R{}_R-T^{\text{EM}}{}^\theta{}_\theta~.
\end{equation}
Given these definitions for the pressures, the conservation law $\nabla_\mu T^{\mu r}=0$ implies
\begin{equation}
\frac{d p_R^{\mathrm{SR}}}{dR}
+
\frac{2}{R}
\left(
p_R^{\mathrm{SR}} - p_\theta^{\mathrm{SR}}
\right)
=
\frac{p_{\mathrm{ext}}(R)}{R}~,
\end{equation}
where $p_{\mathrm{ext}}(R)$ is given by
\begin{equation}
p_{\mathrm{ext}}(R)=-R\left(\frac{d p_R^{\mathrm{LR}}}{dR}
+
\frac{2}{R}
\left(
p_R^{\mathrm{LR}} - p_\theta^{\mathrm{LR}}
\right)
\right)~.
\end{equation}
The external pressure is defined in Ref.~\cite{Farakos:2025byy} as follows
\begin{equation}
P_{\mathrm{ext}}(R)=\frac{1}{R^2}\int_R^\infty dr'\;r'p_{\mathrm{ext}}(r')~,
\end{equation}
whereas the total radial force $\mathcal{F}_{\text{R total}}(r)$ is given by
\begin{equation}
    \mathcal{F}^{\text{SR}}_{R\;\text{total}}(R)=4\pi R^2\left[p^{\text{SR}}_R(R)+P_{\mathrm{ext}}(R)\right]~,\label{radial_force}
\end{equation}
and the other two components of the force are given by
\begin{equation}
    \mathcal{F}^{\text{SR}}_{\theta}(\theta)=2\pi \sin{\theta} \int_{0}^{\infty} dr'\, r' p_{\theta}^{\mathrm{SR}}(r')~,\label{polar_force}
\end{equation}
and
\begin{equation}
    \mathcal{F}^{\text{SR}}_{\phi}=\pi\int_{0}^{\infty} dr'\, r' p_{\theta}^{\mathrm{SR}}(r')~.\label{azimuthal_force}
\end{equation}

Stronger mechanical stability criteria than Laue condition are introduced in Ref.~\cite{Farakos:2025byy}, according which the total radial force $\mathcal{F}^{\text{SR}}_{R\;\text{total}}$ should be positive at radii $R$ close to and inside the monopole core to avoid the collapse of the soliton configurations, while the azimuthal and polar forces should be finite to avoid angular instabilities.   

In the following paragraphs, we examine whether the self-gravitating magnetic monopole and dyon configurations obtained in Sections~\ref{sec:GlobalDilaton} and \ref{sec:GlobalAxion}, respectively, satisfy the energy conditions and the local mechanical stability criteria.

\subsubsection{Energy conditions and mechanical stability for Magnetic Monopoles}
\label{subsec:mon_mech_stab}

\begin{figure}
    \centering
    \includegraphics[width=0.5\textwidth]{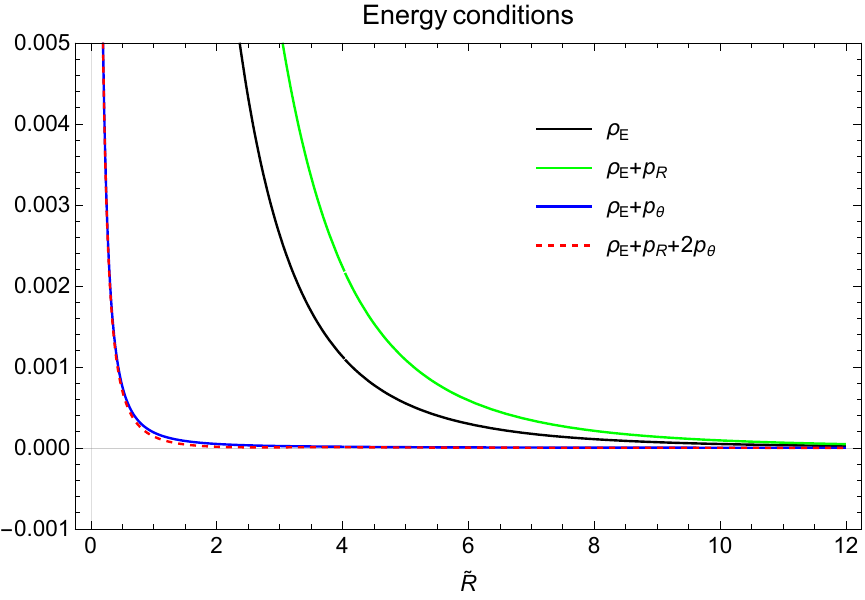}
    \caption{Energy conditions for the numerical magnetic monopole solution with a dilaton charge (Sec.~\ref{sec:GlobalDilaton}), shown as functions of the dimensionless radius $\tilde{R}$ for $\eta=10^{-2}$, $\lambda=10^{4}$, $\beta_{\mathrm{BI}}=1$, and $\zeta=4\sqrt{3}$.}
    \label{fig:6}
\end{figure}

To examine the physical consistency of the magnetic monopole solutions obtained in Sec.~\ref{sec:GlobalDilaton}, we check whether they satisfy standard energy conditions.
In Fig.~\ref{fig:6}, the quantities $\rho_{E}$, $\rho_{E}+p_{R}$, $\rho_{E}+p_{\theta}$, and $\rho_{E}+p_{R}+2p_{\theta}$ are plotted as functions of the dimensionless radius $\tilde{R}$, using the numerical magnetic monopole solution of Sec.~\ref{sec:GlobalDilaton}. As illustrated in Fig.~\ref{fig:6}, all these quantities remain positive throughout spacetime, indicating that all energy conditions are satisfied. These results are consistent with the absence of curvature singularities and event horizons established earlier, as well as further support the physical consistency of the solutions, indicating that the monopole configurations satisfy standard energy requirements while remaining free of singularities and horizons.

\begin{figure}
    \centering
    \includegraphics[width=0.4\textwidth]{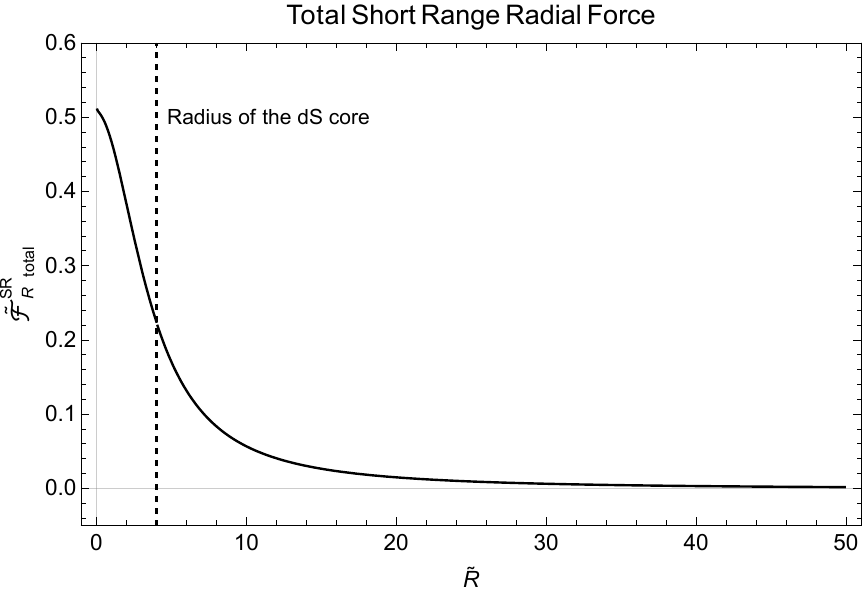}
    \includegraphics[width=0.4\textwidth]{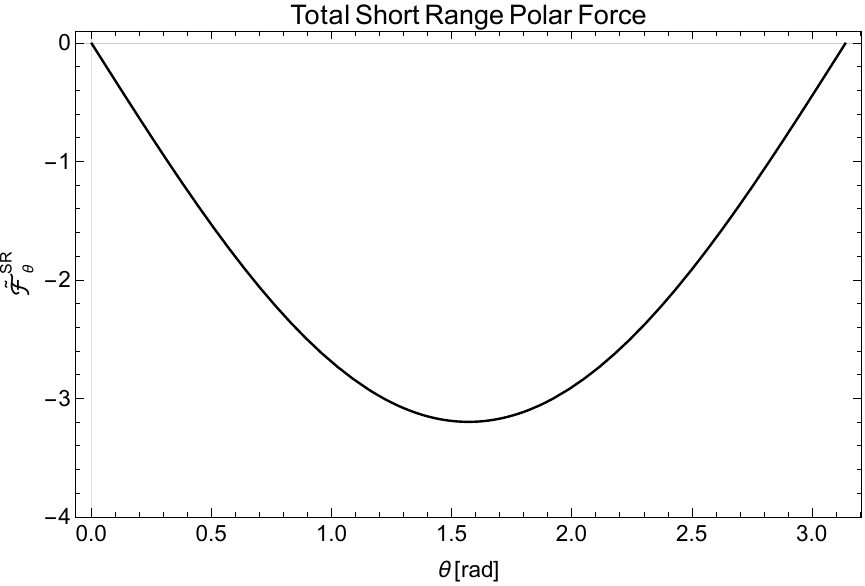}
    \caption{Dimensionless total SR radial (left) and polar (right) forces as functions of the dimensionless radius $\tilde{R}$ and the polar angle $\theta$, respectively, for the numerical magnetic monopole solution obtained in Sec.~\ref{sec:GlobalDilaton} for $\eta=10^{-2}$, $\lambda=10^{4}$, $\beta_{\mathrm{BI}}=1$, and $\zeta=4\sqrt{3}$.}
    \label{fig:Forces_MM}
\end{figure}

In Fig.~\ref{fig:Forces_MM}, we plot the dimensionless total SR radial force 
$\tilde{\mathcal{F}}^{\mathrm{SR}}_{R\,\mathrm{total}}$, defined in 
Eq.~\eqref{radial_force}, and the total SR polar force 
$\tilde{\mathcal{F}}^{\mathrm{SR}}_{\theta}$, defined in 
Eq.~\eqref{polar_force}, in the left and right panels, respectively, as 
functions of the dimensionless radius $\tilde{R}$ and the polar angle 
$\theta$ for the numerical magnetic monopole solution obtained in 
Section~\ref{sec:GlobalDilaton}. 
The total SR radial force remains positive both inside and outside the dS core 
of the monopole, implying that it points outwards and therefore prevents the 
monopole from collapsing. Consequently, the first mechanical stability 
condition, 
$\tilde{\mathcal{F}}^{\mathrm{SR}}_{R\,\mathrm{total}}>0$, is satisfied. 
Furthermore, the polar force remains finite, and therefore the azimuthal force, 
given by Eq.~\eqref{azimuthal_force}, is also finite. As a result, no angular 
instabilities arise, and the angular mechanical stability criteria are likewise 
satisfied.  

At very large radii ($\tilde{R}\to\infty$), the total SR radial force approaches the small negative value 
$\tilde{\mathcal{F}}^{\mathrm{SR}}_{R\,\mathrm{total}}\to -4\pi\eta^2$. 
Hence, the force becomes negative far away from the monopole, in the region where the deficit solid angle contribution proportional to $\eta^2$ dominates. 
Nevertheless, this does not indicate a mechanical instability of the monopole configuration, since the radial SR force remains outward-directed within and near the monopole core.

\subsubsection{Energy conditions and mechanical stability for Dyons}
\label{subsec:dyon_mech_stab}

This subsection extends the purely magnetic analysis of
Section~\ref{sec:mechstab} to the dyonic solutions discussed in
Section~\ref{sec:GlobalAxion}.  We give explicit expressions for the
energy--momentum components, verify the standard energy conditions, and
check local force balance and shell stresses.

For the exterior dyonic background (Eq.~\eqref{EMstrengthAxion}) the
Born-Infeld (BI) electromagnetic stress tensor is
\begin{equation}
  T^{\mu}{}_{\nu}
  =\frac{e^{\Phi}}{4\pi}\!\left[
     \beta_\text{BI}\bigl(1-\sqrt{\Delta}\bigr)\delta^{\mu}{}_{\nu}
    +\frac{e^{-2\Phi}F^{\mu\rho}F_{\nu\rho}
           -\dfrac{e^{-4\Phi}}{2\beta_\text{BI}^{2}}
             (F F^{\!*})F^{\mu\rho}F^{*\!}{}_{\nu\rho}}
          {\sqrt{\Delta}}
    \right],
  \qquad
  \Delta=1+\frac{e^{-2\Phi}X}{2\beta_\text{BI}^{2}}
            -\frac{e^{-4\Phi}Y^{2}}{16\beta_\text{BI}^{4}}.
  \label{eq:dyon_TI}
\end{equation}
Here
\begin{equation}
  X=F^{2}=-\frac{2\!\bigl(Q_m^{2}+{Q_e^{\text{eff}}}^{2}\bigr)}{R^{4}},
  \qquad
  Y=F F^{\!*}=-\frac{4Q_m Q_e^{\text{eff}}}{R^{4}},
  \qquad
  Q_e^{\text{eff}}(r)=Q_e-(b-b_0)Q_m,
  \qquad
  Q^{2}=Q_m^{2}+Q_e^{2}.
\end{equation}
This yields the anisotropic fluid variables
\begin{subequations} \label{eq:dyon_rho_p}
\begin{align}
  \rho_E &=
   \frac{e^{\Phi}}{4\pi}
   \bigl[\sqrt{\beta_\text{BI}^{2}
            +e^{-2\Phi}Q^{2}/R^{4}}
          -\beta_\text{BI}\bigr],
\\
  p_R &=\rho_E
         -\frac{e^{-\Phi}Q^{2}}
                {4\pi R^{4}
                 \sqrt{\beta_\text{BI}^{2}+e^{-2\Phi}Q^{2}/R^{4}}},
\\
  p_\theta &=p_\varphi=
          \rho_E
          +\frac{e^{-\Phi}\bigl(Q_m^{2}-{Q_e^{\text{eff}}}^{2}\bigr)}
                 {4\pi R^{4}
                  \sqrt{\beta_\text{BI}^{2}+e^{-2\Phi}Q^{2}/R^{4}}}.
\end{align}
\end{subequations}

\subsubsection*{Energy conditions}
\begin{itemize}
  \item \textbf{NEC / WEC:}  
        Since $\rho_E>0$ and $\rho_E+p_R>0$ by
        Eqs.~\eqref{eq:dyon_rho_p},
        while $|Q_e^{\text{eff}}|\le Q$ guarantees
        $\rho_E+p_\theta>0$, the null and weak energy conditions are obeyed.
  \item \textbf{SEC:}  
        Adding pressures,
        $\rho_E+p_R+2p_\theta>0$ for all $r\ge R_\text{core}$, so the
        strong energy condition also holds.
\end{itemize}

A numerical scan over $0\le Q_e/Q_m\le1$,
$\tilde\zeta\ge4\sqrt{3}$ and $\beta_\text{BI}\!\gtrsim\!100\,$GeV
can confirm that the minimum of $\rho_E+p_\theta$ stays positive.

\subsubsection*{Mechanical stability}

The effective electric charge of the dyonic solution does not qualitatively alter the properties of the EM stress-energy tensor, while the Kalb--Ramond axion field $b$ contributes positively to the total radial pressure of the anisotropic fluid. In addition, the axion field vanishes at the origin. Consequently, the contributions of the axion field and the effective electric charge to the total SR radial, polar, and azimuthal forces do not qualitatively modify their behavior. Therefore, the dyonic solution also satisfies the mechanical stability conditions discussed for the magnetic monopole solution. 



\subsubsection*{Summary}

To summarize, the Born--Infeld dyon satisfies the energy conditions and is mechanically stable precisely within the parameter region that yields a positive ADM mass and linear perturbative stability, the latter of which is discussed in the following sections. In Appendix~\ref{app:mono_mech_stab}, we further demonstrate that the Laue equilibrium condition is satisfied and that the shell stresses at the surface of the monopole dS core are positive, thereby providing additional confirmation of the stability of our solutions.

\subsection{Linear Dynamical Stability: Formalism}
\label{sec:stabfoprm}

We now apply the GV linear dynamical stability analysis~\cite{Gervalle:2022npx} to the Born--Infeld-regularised monopole model \eqref{model2}. The analysis complements the mechanical stability study above and provides an independent check of stability.

We work with static and spherically symmetric background geometries of the form \eqref{MetricGlobal}. Outside the monopole core the analytic exterior solution \eqref{metricB}, \eqref{dilatonfield} is valid. For the dyon, the axion field is non-trivial in the exterior and the field strength acquires an electric component given in \eqref{EMstrengthAxion}.

The full solution is constructed in a piecewise manner. Inside the core ($r<\delta$) the Higgs field has not reached its symmetry-breaking vacuum, and the energy density is dominated by the approximately constant potential $V\approx\lambda\eta^4/4$, which acts as a positive cosmological constant. The interior metric is therefore approximated by de~Sitter space,
\begin{equation}
A_{\rm int}=B_{\rm int}=1-\frac{2\pi G\lambda\eta^4}{3}r^2.
\label{eq:dScore}
\end{equation}
The shell radius $\delta$ and the mass $M$ are fixed by the Israel matching conditions at $r=\delta$, as derived in detail in the previous sections; positive ADM mass requires $\tilde\zeta>4\sqrt{3}$ in the dimensionless notation of Section~\ref{sec:GlobalDilaton}. For the stability analysis we focus on the exterior region $r>\delta$, where the interior acts as a regular inner boundary condition. Since the interior operator (a Schrödinger operator with positive centrifugal barrier $\ell(\ell+1)/r^2$ plus the de~Sitter curvature contribution $m_{\rm eff}^2 = 2\Lambda_{\rm eff}/3 > 0$) is positive definite, any unstable mode must arise from the exterior, and the interior simply imposes regularity at $r=0$.

\color{black} 
In the current manuscript we shall not study the full problem, that is the study of dilaton, electromagnetic, scalar-triplet and gravity sectors perturbations. We sketch the full analysis of linearized perturbations \eqref{eq:full_linearised_field_content} in Appendix \ref{subsec:epsilon_mixing}, but postpone a complete study to another publication. 

As far as decoupling of dilaton perturbations from the electromagnetic sector are concerned, 
this would happen if the dilaton had been stabilised (possibly through a non-perturbatively-induced potential in string theory) to a classical configuration $\Phi_{\rm cl}(r)$, given, \emph{e.g.}, by 
\eqref{dilatonfield}, or \eqref{dil2}.
If the stabilising potential is such that 
\begin{align}\label{dilsecond}
V^{\prime\prime}(\Phi_{\rm cl}) \gg \omega^2 \,,
\end{align}
where the prime denotes differentiation with respect to $\Phi$ field, and $\omega$ is a characteristic energy (frequency) of the dilaton sector, then the 
perturbation $\delta \Phi$ is massive; it is suppressed in comparison to the electromagnetic-sector perturbations in the string-inspired low-energy (with respect to the string scale) effective field theory framework we adopt. In other words, integrating it out in our effective field theory framework will only produce higher-order subleading corrections.
In the simple case of \eqref{dilatonfield}, this can be seen explicitly, by observing that the logarithmic solution is maintained near the monopole, $m^{\rm eff}_\Phi\, r \ll 1$, where $r$ is the radial coordinate, and $m^{\rm eff}_\Phi$ the effective dilaton ``mass'':
\[ m^{\rm eff \, 2}_\Phi = V^{\prime\prime} (\Phi_{\rm cl}) + \rm background~corrections~.\]
Linearising $\Phi= \Phi_{\rm cl} + \delta \Phi$, we may generically arrive at an equation that couples the dilaton perturbations to the gauge invariant electromagnetic ones $\delta F$:
\[ \Big(\Box - m^{\rm eff\, 2}_\Phi\Big)
= O(\delta F)\,, \]
which implies asymptotic solutions ($m^{\rm eff}_\Phi\, r \gg 1$):
\[ \delta \Phi \sim \frac{e^{-m^{\rm eff}_\Phi \, r}}{r}\,,\]
thereby suppressed ($e^{-m^{\rm eff}_\Phi \, r} \ll1 $, for $m_\Phi \, r \gg 1$) compared to the electromagnetic sector photon perturbations which remain long range 
\[ \delta A_\mu \sim \frac{1}{r}\,.\]
Therefore the background geometry may still carry a sizeable dilaton profile,
but small dynamical dilaton perturbations become negligible far away.
Under such an assumption, the 
presence of a non--constant dilaton does not affect the stability of the system, but rather provides the necessary constraints to induce the MM and dyon solutions of the system within our low-energy effective field theory approach.

We may also assume, for the sake of simplicity in the current manuscript, that the Higgs triplet $\chi^a$, $a=1,2,3$ are extremely massive 
\begin{align}\label{masschi}
m_\chi^2 \, \gg \, \omega^2\,,
\end{align}
where $\omega$ is a typical energy scale of the scalar sector. If condition \eqref{masschi} is satisfied, then, in similar spirit to the dilaton case, any contributions in the stability from the scalar sector will be suppressed by terms of order 
\begin{align}\label{orderomegamchi}
\mathcal O(\frac{\omega^2}{m_\chi^2})\,.
\end{align}

Finally, a metric perturbation will not affect significantly the stability of the MM and dyon configurations if their ADM masses are much smaller than the Planck scale:\footnote{\color{black} We might think that the existence of a de~Sitter core in our monopole solutions, which implies that the radial geodesics experience an outward acceleration (``repulsive'' core), and so the avoidance of catastrophic collapse of matter towards $r=0$, 
would also relate to linear stability, under metric perturbations. Unfortunately, this is not so. Linear stability of a perturbation, say radial, $\delta \mathcal X(t,r) $ is guaranteed if: 
$$ \delta X(r,t) \sim e^{-i\omega \, t}, \quad \rm Im (\omega) <  0\,.$$
This depends on the entire spectrum of the linearized operator, not only on whether gravity is attractive or repulsive. Mechanical stability, on the other hand, is guaranteed by the 
existence of a repulsive de~Sitter core.\color{black}} 
\begin{align}\label{monmassplanck}
   M \ll M_{\rm Pl}\,,
\end{align}
which is the case of MM abd dyons that could be produced at terrestrial current and future colliders, of interest to us here. 

Although we recognize that the above does not fully justify our restriction to the EM sector only for the linear dynamical stability analysis. Nonetheless they set concrete criteria, which, if satisfied by our solution, suggests that the conclusion on the stability under EM perturbation may not be affected by the rest of the sectors. Hence, under these assumptions, the electromagnetic perturbation equations provide the leading contribution to the stability analysis.

We now proceed to perform EM-sector stability diagnostics of our solutions.
\color{black} A naive perturbation of the vector potential $\delta A_\mu$ is not gauge invariant and mixes physical degrees of freedom with pure gauge artefacts. We therefore work throughout with gauge-invariant combinations, following the GV strategy~\cite{Gervalle:2022npx}. Writing the background metric in $2+2$ form,
\begin{equation}
ds^2=g_{ab}(x)\,dx^a dx^b+R^2(x)\,\gamma_{AB}d\theta^A d\theta^B,
\end{equation}
and decomposing the perturbation of the vector potential in scalar and vector spherical harmonics,
\begin{align}
\delta A_a &= u_a(t,r)\,Y_{\ell m}(\theta,\varphi),
\label{eq:dAa}\\
\delta A_A &= u^{(e)}(t,r)\,\hat{D}_A Y_{\ell m}
            +u^{(o)}(t,r)\,\hat{S}_A,
\label{eq:dAA}
\end{align}
the gauge-invariant combinations are
\begin{equation}
E_a:=u_a-\partial_a u^{(e)},\qquad
B:=u^{(o)}.
\label{eq:gauge_inv}
\end{equation}
The perturbation of the field strength in mixed components is then manifestly gauge invariant:
\begin{equation}
\delta F_{aA}=E_a\,\hat{D}_A Y_{\ell m}+(\nabla_a B)\,\hat{S}_A.
\label{eq:dFaA}
\end{equation}
The constraint $\nabla_a(R^2 E^a)=0$ contains no second time derivative and describes no independent propagating degree of freedom; one solves it by introducing an even-parity master field $\Psi_e$ such that $R^2 E^a=\epsilon^{ab}\nabla_b\Psi_e$.

To extract the two physical propagating polarisations we introduce a null tetrad adapted to the background,
\begin{equation}
\{\ell^\mu,\,n^\mu,\,m^\mu,\,\bar{m}^\mu\},\quad
\ell\cdot n=-1,\quad m\cdot\bar{m}=+1,
\end{equation}
with $\ell^\mu$ and $n^\mu$ null in the $(t,r)$ plane. The two transverse propagating modes are
\begin{equation}
\psi_+:=\ell^a m^A\,\delta F_{aA},\qquad
\psi_-:=\ell^a\bar{m}^A\,\delta F_{aA},
\label{eq:psipm_def}
\end{equation}
and, using the standard spin-weighted harmonic projections, one obtains
\begin{align}
\psi_+&=-\frac{\sqrt{\ell(\ell+1)}}{\sqrt{2}R}\,\ell^a(E_a+i\nabla_a B)\,
        {}_{+1}Y_{\ell m},
\label{eq:psi+}\\
\psi_-&=+\frac{\sqrt{\ell(\ell+1)}}{\sqrt{2}R}\,\ell^a(E_a-i\nabla_a B)\,
        {}_{-1}Y_{\ell m}.
\label{eq:psi-}
\end{align}
These are the circularly polarised combinations of the gauge-invariant amplitudes, and are themselves gauge invariant. The full derivation of the projected radial potentials is given in Appendix~\ref{app:projection_monopole_full}.

The electromagnetic sector is characterised by the Lagrangian $\mathcal{L}=\mathcal{L}_{\rm BI}+\mathcal{L}_{bF\tilde{F}}$. The excitation tensor $P^{\mu\nu}:=-2\partial\mathcal{L}/\partial F_{\mu\nu}$ satisfies $\nabla_\mu P^{\mu\nu}=0$, and expanding around a background $\bar{F}_{\mu\nu}$ gives the linearised constitutive relation
\begin{equation}
\delta P^{\mu\nu}=\chi^{\mu\nu}{}_{\rho\sigma}\,\delta F^{\rho\sigma},
\label{eq:linconst}
\end{equation}
where $\chi_{\mu\nu\rho\sigma} = -\frac{2\,\partial^2\mathcal{L}}{\partial F^{\mu\nu}\partial F^{\rho\sigma}}$ is the constitutive tensor evaluated on the background. In Born--Infeld--dilaton--axion theory it depends non-trivially on $\bar{F}_{\mu\nu}$, $\Phi(r)$, and $b(r)$, encoding the response of the nonlinear medium.

 The linearised field equation is
\begin{equation}
\nabla_\mu(\chi^{\mu\nu}{}_{\rho\sigma}\,\delta F^{\rho\sigma})=0,
\label{eq:lineom}
\end{equation}
supplemented by the Bianchi identity $\nabla_{[\mu}\delta F_{\nu\rho]}=0$.

Substituting the helicity variables \eqref{eq:psi+}, \eqref{eq:psi-} 
into the linearised field equation \eqref{eq:lineom}, performing the time-harmonic separation $\psi_\pm(t,r)\propto e^{-i\omega t}\psi_\pm(r)$, the standard procedure of introducing the tortoise coordinate
\begin{equation}
\frac{dr_*}{dr}=\frac{1}{\sqrt{A(r)B(r)}}\,,
\label{eq:tortoise}
\end{equation}
and eliminating first radial-derivative terms, yields the Schrödinger-like system
\begin{equation}
-\frac{d^2}{dr_*^2}\begin{pmatrix}\psi_+\\\psi_-\end{pmatrix}
+\mathbf{V}(r)\begin{pmatrix}\psi_+\\\psi_-\end{pmatrix}
=\omega^2\begin{pmatrix}\psi_+\\\psi_-\end{pmatrix},
\label{eq:master}
\end{equation}
with the potential matrix
\begin{equation}
\mathbf{V}(r)=\begin{pmatrix}V_+(r) & \mathbb{W}(r)\\ \mathbb{W}(r) & V_-(r)\end{pmatrix}.
\label{eq:Vmatrix}
\end{equation}
The quadratic action for the perturbations takes the form
\begin{equation}
S^{(2)}\sim\int dr_*\,\boldsymbol{\Psi}^\dagger\,\mathcal{W}(r)
\left[-\frac{d^2}{dr_*^2}+\mathbf{V}(r)\right]\boldsymbol{\Psi},
\label{eq:quad_action}
\end{equation}
where $\mathcal{W}(r)\propto\kappa_\perp(r)\,\mathbf{1}$ is the weight matrix determined by the constitutive tensor. Since $\bar{L}_X(r)=-\bar\Delta(r)^{-1/2}$ with $\bar\Delta(r)>0$ everywhere (the BI discriminant condition is satisfied for all solutions considered here), and the geometric prefactors $R^2(r)>0$ and $\sqrt{A(r)B(r)}>0$ hold in the exterior, $\mathcal{W}(r)$ is positive definite throughout. The inner product
\begin{equation}
\langle\boldsymbol{\Psi}_1,\boldsymbol{\Psi}_2\rangle
=\int dr_*\,\boldsymbol{\Psi}_1^\dagger\mathcal{W}(r)\boldsymbol{\Psi}_2
\label{eq:inner}
\end{equation}
makes the operator $\mathcal{O}=-d^2/dr_*^2+\mathbf{V}(r)$ self-adjoint whenever $\mathbf{V}(r)$ is real and symmetric and $\mathcal{W}(r)$ is positive definite. For static backgrounds with real field profiles---as holds for both the monopole and dyonic solutions---all entries of $\mathbf{V}(r)$ are real, and spherical symmetry implies $V_+(r)=V_-(r)\equiv V_0(r)$. The spectrum $\{\omega^2\}$ is therefore entirely real, and exponential growth is impossible.

\subsection{Linear Stability Analysis of Monopole and Dyon}
\label{sec:stabanal}

We now specialise the formalism to the two explicit solutions.

\subsubsection{Magnetic monopole: decoupled helicities and positive potential}
\label{sec:monopole}

For the purely magnetic monopole the only non-zero background field-strength component is $\bar{F}_{\theta\varphi}=Q_m\sin\theta$, so the pseudoscalar invariant vanishes identically on the background:
\[
\bar Y = \bar F_{\mu\nu}\tilde{\bar F}^{\mu\nu} = 0.
\]
Since the off-diagonal term $\mathbb{W}(r)$ in \eqref{eq:Vmatrix} originates from the $Y$-dependent part of $\chiT$, it vanishes:
\begin{equation}
\mathbb{W}(r)=0\quad\text{(magnetic monopole)}.
\label{eq:Wzero}
\end{equation}
The two helicity sectors decouple and each satisfies an independent Regge--Wheeler-type equation,
\begin{equation}
-\frac{d^2\psi}{dr_*^2}+V_0(r)\psi=\omega^2\psi,
\label{eq:RW}
\end{equation}
where $\psi$ stands for either $\psi_+$ or $\psi_-$.

In the exterior region the dilaton profile is $\Phi(r)=-\ln(1-\zeta/r)$, the areal radius is $R(r)=\sqrt{r(r-\zeta)}$, and the metric function is $A=B=1-8\pi G\eta^2-2GM/r$. The effective potential takes the form
\begin{equation}
V_0(r)=A(r)\!\left[\frac{\ell(\ell+1)}{R^2(r)}
+\frac{V_{\rm BI}(r)+V_{\rm dil}(r)}{R^2(r)}\right],
\label{eq:V0explicit}
\end{equation}
where $V_{\rm BI}(r)$ contains the Born--Infeld nonlinearity corrections and $V_{\rm dil}(r)$ encodes the dilaton coupling. Near the core ($r\to\zeta^+$), since $R^2(r)=r(r-\zeta)\sim\zeta(r-\zeta)\to 0$ while $A(r)$ remains finite and positive, the angular-momentum barrier dominates and one obtains
\begin{equation}
V_0(r)\sim\frac{c_0}{(r-\zeta)},\quad r\to\zeta^+,\quad c_0>0,
\label{eq:V_core}
\end{equation}
a repulsive barrier absent in ordinary Schwarzschild or Reissner--Nordström backgrounds. Asymptotically ($r\to\infty$), with $R(r)\sim r$ and $A(r)\to1-8\pi G\eta^2$,
\begin{equation}
V_0(r)\sim(1-8\pi G\eta^2)\frac{\ell(\ell+1)}{r^2}+O(r^{-3}).
\label{eq:V_infty}
\end{equation}
Both limits are manifestly positive. The numerical solutions of Section~\ref{sec:GlobalDilaton} confirm that $A(r)=B(r)>0$ on the entire exterior domain (no horizon forms for the physically acceptable branch $\tilde\zeta\ge4\sqrt{3}$), so the angular-barrier term $A(r)\,\ell(\ell+1)/R^2(r)$ is strictly positive for $\ell\ge 1$ throughout the exterior. The Born--Infeld correction $V_{\rm BI}(r)$ is non-negative because it arises from $\bar{L}_{XX}>0$. The dilaton correction $V_{\rm dil}(r)$ may change sign, but its magnitude is suppressed relative to the angular barrier by a factor of order $Q_m^2/(\beta_{\rm BI}\,r^2)$, which is small in the exterior. Hence $V_0(r)>0$ on the full domain $r\in(\zeta,\infty)$.

Note that for $\ell=0$ the spin-weighted harmonics ${}_{s}Y_{0m}$ vanish identically for $|s|\ge 1$, so the helicity variables $\psi_\pm$ vanish for $\ell=0$: there are no propagating electromagnetic degrees of freedom in this sector, only the Gauss-law constraint fixing the total magnetic charge. The stability analysis is therefore vacuous for $\ell=0$, and all statements about $V_0>0$ apply for $\ell\ge 1$.

Because $V_0(r)>0$ everywhere in the exterior and the operator $\mathcal{O}=-d^2/dr_*^2+V_0$ is self-adjoint with respect to \eqref{eq:inner}, its spectrum satisfies
\begin{equation}
\omega^2=\frac{\langle\psi,\mathcal{O}\psi\rangle}{\langle\psi,\psi\rangle}
=\frac{\int dr_*\,|\psi'|^2+\int dr_*\,V_0|\psi|^2}{\int dr_*\,|\psi|^2}>0,
\label{eq:omega2_pos}
\end{equation}
for any normalisable perturbation. There are no modes with $\omega^2<0$, and the purely magnetic self-gravitating Born--Infeld monopole is therefore \textbf{linearly mode stable}. The repulsive near-core barrier \eqref{eq:V_core} provides an additional layer of stability, reflecting perturbations away from the core region by a potential wall that grows without bound as $r\to\zeta^+$.

\subsubsection{Dyon: helicity mixing and the $2\times 2$ Sturm--Liouville problem}
\label{sec:dyon}

For the dyonic background \eqref{EMstrengthAxion}, both electric and magnetic fields are present and the pseudoscalar invariant $\bar Y = \bar F_{\mu\nu}\tilde{\bar F}^{\mu\nu}$ is non-vanishing:
\begin{align}\label{psscinv}
\bar Y\propto \frac{Q_m\,Q_e^{\rm eff}(r)}{R^4(r)} \ne 0\,.
\end{align}
The constitutive tensor $\chiT$ therefore acquires an off-diagonal structure in the helicity basis: the term $\partial^2\mathcal{L}/\partial F_{\mu\nu}\partial F_{\rho\sigma}$ evaluated on the dyonic background couples $\delta F$ components of opposite helicity through the background $\tilde{\bar{F}}$ tensor. This is the mechanism identified in the GV framework~\cite{Gervalle:2022npx}: when $\bar{Y}\neq 0$, the constitutive tensor mixes the two helicity sectors.

The off-diagonal coupling arises from the part of the constitutive tensor that is odd in the pseudoscalar invariant \eqref{psscinv}.  After projection onto the helicity basis, the off-diagonal entry of the potential matrix takes the explicit form
\begin{equation}
\mathbb{W}(r)=-\frac{4\,C_\ell(r)}{\beta\,\Delta^{3/2}(r)}\cdot
\frac{e^{-2\Phi(r)}\,Q_m\,\Qeff(r)}{R^4(r)},
\label{eq:Wexplicit}
\end{equation}
where $C_\ell(r)=\ell(\ell+1)/(2R^2(r))$ is the angular projection coefficient (exact from the spin-weighted harmonic identities), and $\Delta(r)$ is the Born--Infeld discriminant evaluated on the background. The key structural feature is
\begin{equation}
\mathbb{W}(r)\propto Q_m\bigl[Q_e-(b(r)-b_0)Q_m\bigr],
\label{eq:W_structure}
\end{equation}
so that in the purely magnetic limit ($Q_e=0$, $b(r)=b_0$) one has $\Qeff=0$ and $\mathbb{W}(r)=0$, recovering \eqref{eq:Wzero}. More generally, the axion field $b(r)$ partially screens the electric charge.

The full dyonic perturbation equation is the $2\times 2$ matrix system \eqref{eq:master} with
\begin{equation}
\mathbf{V}(r)=V_0(r)\,\mathbf{1}+\mathbb{W}(r)\,\sigma_x,
\label{eq:Vdyon}
\end{equation}
where $\sigma_x=\begin{pmatrix}0&1\\1&0\end{pmatrix}$. Since the static dyonic field profiles $\Phi(r)$, $b(r)$ and the metric functions are all real, $V_0(r)$ and $\mathbb{W}(r)$ are real functions and $\mathbf{V}(r)$ is a real symmetric matrix. The operator $\mathcal{O}=-d^2/dr_*^2\,\mathbf{1}+\mathbf{V}(r)$ is self-adjoint and its eigenvalues $\omega^2$ are real.

To establish positivity, diagonalise $\sigma_x$ by $U=\frac{1}{\sqrt{2}}\begin{pmatrix}1&1\\1&-1\end{pmatrix}$:
\begin{equation}
U\mathbf{V}U^\dagger=\begin{pmatrix}V_0+\mathbb{W} & 0\\0 & V_0-\mathbb{W}\end{pmatrix}.
\label{eq:diag}
\end{equation}
Stability requires both eigenvalues $V_0\pm\mathbb{W}$ to be positive. The ratio $|\mathbb{W}(r)|/V_0(r)$ satisfies
\begin{equation}
\frac{|\mathbb{W}(r)|}{V_0(r)}
\le
\frac{2\,e^{-2\Phi(r)}\,|Q_m\,\Qeff(r)|}
     {\beta\,\Delta^{3/2}(r)\,A(r)\,R^4(r)}.
\label{eq:WV0ratio}
\end{equation}
Asymptotically ($r\to\infty$): since $\Qeff(r)\to Q_e-(b(\infty)-b_0)Q_m=0$ by the boundary condition $b_0=-Q_e/Q_m$, the ratio vanishes. Near the core ($r\to\zeta^+$): $\Delta(r)\to\infty$ because $\bar X\sim R^{-4}\to\infty$, so $\Delta^{3/2}$ suppresses $\mathbb{W}(r)$ faster than $V_0(r)$ diverges---concretely, $|\mathbb{W}|\sim(r-\zeta)^{+1/2}$ while $V_0\sim(r-\zeta)^{-1}$, so $|\mathbb{W}|/V_0\to 0$ as $r\to\zeta^+$. Hence $|\mathbb{W}(r)|<V_0(r)$ throughout the exterior, both eigenvalues $V_0\pm\mathbb{W}$ are positive definite, and $\omega^2>0$ for all normalisable modes. The dyonic solutions are therefore \textbf{linearly mode stable}.

The helicity mixing does not introduce instabilities but rather produces a birefringent polarisation structure. In the monopole case the effective medium is isotropic and propagation is conformally equivalent to propagation in the background geometry. In the dyonic case, $\bar{Y}\neq 0$ breaks this symmetry: the two helicities see different effective metrics,
\begin{equation}
g^{\mu\nu}_{\rm eff}(\pm)\propto g^{\mu\nu}\pm\Delta g^{\mu\nu}(\bar{F},\tilde{\bar{F}}),
\end{equation}
an effect analogous to birefringence in anisotropic optical media. The off-diagonal term $\mathbb{W}(r)$ quantifies this birefringence and vanishes precisely when $\Qeff=0$, i.e.\ when the axion screens the electric charge completely.

\section{Discussion and conclusions}
\label{sec:conclude}

In this work, we have shown that global monopoles coupled to string-inspired dilaton, axion, and Born--Infeld (BI) sectors admit physically viable, self-gravitating magnetic monopole and dyon solutions. In particular, we constructed magnetically charged monopole configurations and their dyonic generalisations, demonstrating that the inclusion of dilaton and axion couplings converts the negative-mass behaviour of the standard global monopole into solutions with positive ADM mass. The resulting configurations are free of curvature singularities and event horizons, possess finite self-energy due to the non-linear BI electrodynamics, and satisfy the standard energy conditions. Furthermore, a linear perturbation analysis of the electromagnetic sector indicates that both the monopole and dyon solutions are dynamically stable, with no evidence of unstable modes despite the presence of axion-induced helicity mixing in the dyonic case.

In particular, we constructed magnetically charged monopole configurations in which the dilaton field generates a secondary scalar hair, linking the magnetic charge to both the dilaton charge and the monopole mass. By employing an approximate analytic treatment according which the monopole possesses a de Sitter core, we imposed the Israel conditions to match the interior de Sitter spacetime to an exterior geometry, and identified the parameter regime in which the monopole mass becomes positive. Although this approximation is not uniformly accurate across the entire parameter space, our numerical analysis confirmed the existence of regular solutions and showed that the analytic approximate predictions capture the correct order of magnitude for both the mass and the magnetic charge.

A notable feature of the model is that the Born--Infeld sector renders the monopole self-energy finite, in contrast to the divergent behaviour encountered in Maxwell electrodynamics. This provides an additional indication that non-linear electrodynamics offers a natural framework for regulating classical field configurations with point-like sources.
We further extended the analysis by incorporating a KR axion field, leading to dyonic generalizations of the monopole solutions. In this case, the axion coupling induces an effective electric charge through its interaction with the electromagnetic sector, resulting in a non-trivial dressing of the electric field. The resulting configurations are characterized by a combined charge $Q^{2}=Q_{e}^{2}+Q_{m}^{2}$, while the interior structure remains essentially unchanged under appropriate boundary conditions.

We also examined the mechanical stability and energy conditions of the solutions. Using the numerical configurations (in the monopole case), we showed that the energy density and pressures satisfy the null, weak, and strong energy conditions throughout spacetime. This behaviour is consistent with the absence of curvature singularities and horizons, and supports the physical viability of the solutions.

In addition, we performed a systematic linear mode stability analysis of the self-gravitating
Born--Infeld monopole and dyonic solutions. We recall the key steps:
\begin{enumerate}
\item Starting from the perturbation of the vector potential $\delta A_\mu$, we
  constructed the gauge-invariant amplitudes $(E_a,B)$ via the standard harmonic
  decomposition~\eqref{eq:gauge_inv}.
\item Projecting onto a null tetrad, we identified the two helicity modes $\psi_\pm$
  as the physically propagating degrees of freedom~\eqref{eq:psi+}--\eqref{eq:psi-}.
\item The linearised constitutive-tensor equation~\eqref{eq:lineom} reduces, after
  angular separation and introduction of the tortoise coordinate, to the
  Schr\"{o}dinger-like master equation~\eqref{eq:master} with a real symmetric
  potential matrix.
\item Self-adjointness of the radial operator immediately implies a real spectrum
  $\omega^2\in\mathbb{R}$.
\item Positivity of the effective potential $V_0(r)>0$ in the exterior, together with
  the rigorous bound $|\mathbb{W}(r)|<V_0(r)$ derived from the explicit form
  $C_\ell(r)=\ell(\ell+1)/(2R^2(r))$ and the BI saturation near the core,
  establishes $\omega^2>0$ and hence linear stability for both monopoles and dyons.
\end{enumerate}

For the magnetic monopole the analysis is particularly transparent: the two helicity sectors
are completely decoupled, the effective potential is positive definite with a repulsive
near-core barrier $V_0\sim(r-\zeta)^{-1}$ arising from the dilaton structure of the
exterior geometry, and linear stability follows directly from the positivity
argument~\eqref{eq:omega2_pos}.

For the dyon, the axion-induced $F\tilde{F}$ coupling generates a non-trivial
off-diagonal mixing term $\mathbb{W}(r)\propto Q_m\Qeff(r)$.  The angular coefficient
$C_\ell(r)=\ell(\ell+1)/(2R^2(r))$ is computed from the spin-weighted harmonic
projection, and the bound~\eqref{eq:WV0ratio} shows rigorously that
$|\mathbb{W}(r)|/V_0(r)\to 0$ both asymptotically (because $\Qeff(r)\to 0$ by the boundary
condition $b_0=-Q_e/Q_m$) and near the core (because $\Delta^{3/2}\to\infty$
suppresses $\mathbb{W}$ faster than $V_0$ diverges).  Hence both eigenvalues $V_0\pm W$
are positive throughout, confirming stability without any additional assumption.
The helicity mixing does not introduce instabilities but rather produces a
birefringent polarisation structure in which the two circular polarisations propagate
in slightly different effective geometries.

A notable structural result is the existence of a minimum nonzero magnetic charge
in this model: as the de~Sitter core radius approaches its lower bound
$\tilde{R}_{\mathrm{dS}}\to\sqrt{6}$, the monopole mass vanishes while the magnetic
charge $\tilde{Q}_m=\sqrt{\tilde{\zeta}\tilde{M}}$ remains finite, so the
zero-mass configuration is held up entirely by magnetic charge and de~Sitter core
pressure.
Overall, our results indicate that string-inspired dilaton, axion, and Born--Infeld couplings provide a natural framework in which global monopoles can be promoted to physically viable, self-gravitating magnetic monopoles and dyons with positive mass, finite energy, and stable dynamics. 

Several directions for future work remain open. The gravitational sector perturbations
(metric and dilaton fluctuations) have not been treated here; a full stability analysis
would require coupling the electromagnetic GV system to the Regge--Wheeler--Zerilli
equations for the metric and a coupled scalar equation for the dilaton.  For the
dyonic case, incorporating the gravitational back-reaction on the helicity-mixing
structure could modify the effective potentials at sub-leading order \cb (see Appendix \ref{subsec:epsilon_mixing}) \cbl.   
Finally, it would be of interest to explore the phenomenological implications of these configurations, in particular their potential observational signatures in gravitational or electromagnetic contexts. Indeed, given the absence of compositeness in our MM, since our solutions are not characterised by constituent gauge bosons, 
in contrast to what happens in the electroweak~\cite{Cho:1996qd,Cho:2013vba,Ellis:2016glu} or low-mass Grand Unified MM~\cite{Kephart:2017esj}, but they rather approximate Dirac-type structureless MM, 
one might hope of avoiding the strong suppression for the production of such MM at colliders, provided concrete relevant processes are identified. This will be looked at in future works. Moreover, such configurations may characterise phase transitions in the post-inflationary early Universe, and may have significant astrophysical or cosmological signatures, especially as a result of the interactions  of the self-gravitating MM with gravitational waves or cosmic microwave background radiation. 

\emph{Affaire \`a suivre ...}

\section*{Acknowledgments}
N.E.M.\ thanks the University of Valencia and its Theoretical Physics Department for
a visiting Research Professorship supported by the programme Atracci\'{o}n de Talento
INV25-01-15, during which this work has started.  The work of N.E.M.\ and S.S.\ is supported in part by 
the Science and Technology Facilities
Council (STFC) under the research grant no. ST/X000753/1,
respectively.  N.E.M.\ also acknowledges participation in the COST Association Actions
CA21136 ``Addressing observational tensions in cosmology with systematics and
fundamental physics (CosmoVerse)'' and CA23130 ``Bridging high and low energies in
search of Quantum Gravity (BridgeQG)''. 
D. T. acknowledges support by the U.S. Department of Energy, Office of Science, Office of High Energy Physics program under
Award No. DE-SC-0022021.

\appendix

\section{Projection of the perturbation equations and derivation of the monopole radial potentials}
\label{app:projection_monopole_full}

This appendix gives a self-contained derivation of the radial Sturm--Liouville problem governing electromagnetic perturbations of the purely magnetic monopole background, making precise the origin of the structural contributions $V_{\rm BI}(r)$ and $V_{\rm dil}(r)$ that appear in the main text. Throughout we restrict to the purely magnetic case,
\begin{equation}
\bar F_{\theta\phi}=Q_m \sin\theta,
\qquad
\bar F_{tr}=0,
\qquad
\bar Y:=\bar F_{\mu\nu}\tilde{\bar F}^{\mu\nu}=0,
\label{eq:monopole_background_appendix}
\end{equation}
so that the two helicity sectors decouple from the outset.

\subsection{Linearised constitutive equation and Born--Infeld coefficients}

The electromagnetic equations of motion $\nabla_\mu P^{\mu\nu}=0$, with $P^{\mu\nu}:=-2\partial\mathcal L/\partial F_{\mu\nu}$, are supplemented by the Bianchi identity $\nabla_{[\mu}F_{\nu\rho]}=0$. Perturbing around the background, $F_{\mu\nu}=\bar F_{\mu\nu}+\delta F_{\mu\nu}$, the linearised relation is
\begin{equation}
\delta P^{\mu\nu}
=
\chi^{\mu\nu}{}_{\rho\sigma}\,\delta F^{\rho\sigma},
\qquad
\chi^{\mu\nu}{}_{\rho\sigma}
=
-2\,\frac{\partial^2\mathcal L}{\partial F_{\mu\nu}\partial F^{\rho\sigma}},
\label{eq:chi_def_appendix}
\end{equation}
and the linearised field equation is $\nabla_\mu(\chi^{\mu\nu}{}_{\rho\sigma}\,\delta F^{\rho\sigma})=0$.

The Born--Infeld Lagrangian depends on the two invariants $X:=F_{\mu\nu}F^{\mu\nu}$ and $Y:=F_{\mu\nu}\tilde F^{\mu\nu}$ through
\begin{equation}
\mathcal L_{\rm BI}
=
4\beta^2 e^{2\Phi}
\left(
1-\sqrt{\Delta}
\right),
\qquad
\Delta
:=
1+\frac{e^{-2\Phi}}{2\beta^2}X
-\frac{e^{-4\Phi}}{16\beta^4}Y^2.
\label{eq:LBI_appendix}
\end{equation}
The relevant derivatives are
\begin{equation}
L_X := \frac{\partial\mathcal L}{\partial X} = -\frac{1}{\sqrt{\Delta}},
\qquad
L_{XX} = \frac{e^{-2\Phi}}{4\beta^2}\,\Delta^{-3/2}.
\label{eq:LX_appendix}
\end{equation}
Since $\bar Y=0$ for the purely magnetic monopole, the background dependence enters only through
\begin{equation}
\bar X(r)=\frac{2Q_m^2}{R^4(r)},
\qquad
\bar\Delta(r)=1+\frac{e^{-2\Phi(r)}}{2\beta^2}\bar X(r),
\label{eq:Xbar_Deltabar_appendix}
\end{equation}
giving
\begin{equation}
\bar L_X(r)=-\bar\Delta(r)^{-1/2},
\qquad
\bar L_{XX}(r)=\frac{e^{-2\Phi(r)}}{4\beta^2}\bar\Delta(r)^{-3/2}.
\label{eq:LXbar_LXXbar_appendix}
\end{equation}

\subsection{Gauge-invariant variables and helicity projection}

We write the metric in $2+2$ form, $ds^2=g_{ab}(x)\,dx^a dx^b + R^2(x)\gamma_{AB}\,d\theta^A d\theta^B$ with $x^a=(t,r)$, and decompose the vector-potential perturbation in scalar and vector spherical harmonics as in equations~\eqref{eq:dAa}--\eqref{eq:dAA} of the main text. The gauge-invariant amplitudes are $E_a:=u_a-\partial_a u^{(e)}$ and $B:=u^{(o)}$, and the mixed field-strength perturbation is
\begin{equation}
\delta F_{aA}
=
E_a\,\hat D_A Y_{\ell m}
+
(\nabla_a B)\,\hat S_A.
\label{eq:dFaA_appendix}
\end{equation}

To extract the propagating helicity modes we introduce a complex null tetrad $\{\ell^\mu, n^\mu, m^\mu, \bar m^\mu\}$ with $\ell\cdot n=-1$, $m\cdot\bar m=+1$, all other inner products vanishing. The tetrad is adapted to the $2+2$ split: $\ell^\mu$ and $n^\mu$ are purely radial, $m^\mu=(0,m^A)$ and $\bar m^\mu=(0,\bar m^A)$ are purely angular, with $m^A$ and $\bar m^A$ forming a complex null basis on $S^2$ satisfying $\gamma_{AB}m^A\bar m^B=1$. The area form acts on them as $\hat\epsilon_A{}^B m_B=+im_A$ and $\hat\epsilon_A{}^B\bar m_B=-i\bar m_A$, so they are eigenvectors of the Hodge dual on $S^2$ with eigenvalues $\pm i$. The helicity variables are
\begin{equation}
\psi_+:=\ell^a m^A\,\delta F_{aA},\qquad
\psi_-:=\ell^a\bar m^A\,\delta F_{aA}.
\label{eq:psi_helicity_appendix}
\end{equation}
Using \eqref{eq:dFaA_appendix} together with the spin-weighted harmonic projections $m^A\hat D_A Y_{\ell m}=-\sqrt{\ell(\ell+1)}/(\sqrt2 R)\,{}_{+1}Y_{\ell m}$ and $m^A\hat S_A=-i\sqrt{\ell(\ell+1)}/(\sqrt2 R)\,{}_{+1}Y_{\ell m}$ (and their conjugates for $\bar m^A$), one obtains
\begin{equation}
\psi_+
=
-\frac{\sqrt{\ell(\ell+1)}}{\sqrt2\,R}\,
\ell^a(E_a+i\nabla_a B)\,{}_{+1}Y_{\ell m},
\qquad
\psi_-
=
+\frac{\sqrt{\ell(\ell+1)}}{\sqrt2\,R}\,
\ell^a(E_a-i\nabla_a B)\,{}_{-1}Y_{\ell m}.
\label{eq:psi_pm_appendix}
\end{equation}
Because $\bar Y=0$, the constitutive tensor has no parity-odd dual-mixing term, so the off-diagonal helicity coupling vanishes ($W(r)=0$) and both helicities satisfy the same scalar equation. Dropping the helicity label,
\begin{equation}
-\frac{d^2\psi}{dr_*^2}+V_0(r)\psi=\omega^2\psi.
\label{eq:monopole_master_appendix}
\end{equation}

\subsection{From weighted Sturm--Liouville to Schr\"{o}dinger form}

Before removing first-derivative terms, the projected radial equation takes the weighted Sturm--Liouville form
\begin{equation}
-\frac{d}{dr}\left(P(r)\frac{d\psi}{dr}\right)
+Q(r)\psi
=
\omega^2 W(r)\psi,
\label{eq:weighted_SL_appendix}
\end{equation}
where the kinetic coefficient is determined by the transverse constitutive response, $P(r)\sim W(r)\sim -\bar L_X(r)$, with geometric prefactors from the $2+2$ decomposition. The potential $Q(r)$ receives three contributions: (i) the angular barrier $\propto\ell(\ell+1)$; (ii) terms from $\bar L_X$ and $\bar L_{XX}$ via the Born--Infeld constitutive tensor; (iii) terms from derivatives of $A(r)$, $B(r)$, $R(r)$, $\Phi(r)$ from the background geometry. This motivates the split $Q(r)=Q_{\rm ang}+Q_{\rm BI}+Q_{\rm dil}$.

To reduce \eqref{eq:weighted_SL_appendix} to Schrödinger form one introduces the tortoise coordinate
\begin{equation}
\frac{dr_*}{dr}=\sqrt{\frac{W(r)}{P(r)}}
\label{eq:drstar_general_appendix}
\end{equation}
and rescales $\psi(r)=[P(r)W(r)]^{-1/4}\,u(r)$, obtaining
\begin{equation}
-\frac{d^2u}{dr_*^2}+V_{\rm eff}(r)\,u=\omega^2 u.
\label{eq:Schrodinger_general_appendix}
\end{equation}
In the case $P(r)=W(r)$ this simplifies to
\begin{equation}
V_{\rm eff}(r)
=
\frac{Q(r)}{P(r)}
+\frac{1}{2}\frac{P''(r)}{P(r)}
-\frac{1}{4}\left(\frac{P'(r)}{P(r)}\right)^2,
\label{eq:Veff_PeqW_appendix}
\end{equation}
which makes the Born--Infeld contribution explicit: any radial variation of $\bar L_X(r)$ generates additional terms through the derivatives of $P(r)$. After the Liouville transformation one writes
\begin{equation}
V_0(r)
=
A(r)\left[
\frac{\ell(\ell+1)}{R^2(r)}
+\frac{V_{\rm BI}(r)+V_{\rm dil}(r)}{R^2(r)}
\right],
\label{eq:V0_split_appendix}
\end{equation}
where $V_{\rm BI}$ collects the terms whose coefficients originate from $\bar L_X$, $\bar L_{XX}$, and $\bar X(r)=2Q_m^2/R^4(r)$, while $V_{\rm dil}$ collects the terms generated by the dilaton-dependent metric functions $R(r)$, $\Phi(r)$, $A(r)$, $B(r)$ and their derivatives.

\subsection{Asymptotic behaviour of the effective potential}

The analytic exterior background is
\begin{equation}
A(r)=B(r)=1-8\pi G\eta^2-\frac{2GM}{r},\quad
R(r)=\sqrt{r(r-\zeta)},\quad
\Phi(r)=-\ln\!\left(1-\frac{\zeta}{r}\right),
\label{eq:exterior_background_appendix}
\end{equation}
giving $\bar X(r)=2Q_m^2/[r^2(r-\zeta)^2]$. Near the inner boundary $r\to\zeta^+$, one has $R^2(r)\sim\zeta(r-\zeta)\to 0$ while $A(r)$ remains finite and positive, so the angular barrier dominates:
\begin{equation}
V_0(r)\sim
\frac{A(\zeta)\,\ell(\ell+1)}{\zeta(r-\zeta)}
\equiv
\frac{c_0}{r-\zeta},
\qquad
c_0>0,
\label{eq:V0_nearzeta_appendix}
\end{equation}
a repulsive barrier that diverges at the dilaton singularity. Asymptotically, $R(r)\sim r$ and $A(r)\to1-8\pi G\eta^2$, so
\begin{equation}
V_0(r)
=
(1-8\pi G\eta^2)\frac{\ell(\ell+1)}{r^2}
+\mathcal O(r^{-3}).
\label{eq:V0_infty_appendix}
\end{equation}
Inside the de~Sitter core, $A_{\rm int}(r)=B_{\rm int}(r)=1-\Lambda_{\rm eff}r^2/3$ with $\Lambda_{\rm eff}=2\pi G\lambda\eta^4$, and near $r=0$,
\begin{equation}
V_{\rm int}(r)
=
\frac{\ell(\ell+1)}{r^2}
+\mathcal O(1),
\qquad r\to0,
\label{eq:Vint_origin_appendix}
\end{equation}
where the $\mathcal O(1)$ term contains the positive de~Sitter curvature correction $m_{\rm eff}^2=2\Lambda_{\rm eff}/3$. Both asymptotic limits of $V_0$ are manifestly positive, and together with the positivity argument established in Section~\ref{sec:monopole} they confirm linear stability without requiring explicit closed-form expressions for $V_{\rm BI}$ and $V_{\rm dil}$.

\cb
\section{Bookkeeping parameter for sector mixing}
\label{subsec:epsilon_mixing}

To clarify the status of the electromagnetic perturbation analysis, it is
useful to introduce a formal bookkeeping parameter $\epsilon_{\rm mix}$
multiplying the couplings between the electromagnetic helicity sector and the
remaining perturbations. The full perturbation vector is schematically
\begin{equation}
  \boldsymbol{\Psi}
  =
  \begin{pmatrix}
    \Psi_{\rm EM}\\[4pt]
    \Psi_{\rm rest}
  \end{pmatrix},
  \qquad
  \Psi_{\rm EM}
  =
  \begin{pmatrix}
    \psi_+\\[4pt]
    \psi_-
  \end{pmatrix},
  \qquad
  \Psi_{\rm rest}
  =
  \begin{pmatrix}
    \varphi\\[2pt]
    \beta\\[2pt]
    H\\[2pt]
    Z_{\rm grav}^{\rm odd}\\[2pt]
    Z_{\rm grav}^{\rm even}
  \end{pmatrix},
  \label{eq:epsilon_mixing_vector}
\end{equation}
where $\varphi=\delta\Phi$, $\beta=\delta b$, $H=\delta h$, and 
$Z_{\rm grav}^{\rm odd}$ and $Z_{\rm grav}^{\rm even}$ denote the
gravitational master variables obtained after decomposing $h_{\mu\nu}$ into
odd- and even-parity spherical harmonics, fixing the metric gauge and
eliminating the non-dynamical Einstein constraint variables. In a vacuum
Schwarzschild background these reduce to the standard Regge--Wheeler and
Zerilli variables, respectively. In the present self-gravitating
monopole/dyon background they should instead be understood as
Regge--Wheeler- and Zerilli-type variables adapted to a non-vacuum spherical
geometry and coupled to the matter perturbations.

We now indicate how the non-electromagnetic components of
$\Psi_{\rm rest}$ enter the full perturbation problem. The purpose of
this appendix is not to carry out the complete Einstein--matter spectral
analysis, but to make explicit which four-dimensional equations are
being linearised and how they fit into the block operator
$\mathcal{L}_{\rm rest}$ appearing in Eq. \eqref{eq:epsilon_mixing_operator}.
 
The full set of perturbations is
\begin{equation}
  g_{\mu\nu}
  =
  \bar{g}_{\mu\nu} + \epsilon\,h_{\mu\nu},
  \qquad
  F_{\mu\nu}
  =
  \bar{F}_{\mu\nu} + \epsilon\,f_{\mu\nu},
  \qquad
  \Phi = \bar{\Phi} + \epsilon\,\varphi,
  \qquad
  b = \bar{b} + \epsilon\,\beta,
  \qquad
  \chi^a = \bar{\chi}^a + \epsilon\,\delta\chi^a,
\label{eq:full_linearised_field_content}
\end{equation}
where barred quantities denote the background monopole or dyon solution.
For the Higgs triplet we write
\begin{equation}
  \bar{\chi}^a
  =
  \eta\,\bar{h}(r)\,\frac{x^a}{r},
  \qquad
  \delta\chi^a
  =
  \eta\,H(t,r,\theta,\phi)\,\frac{x^a}{r}
  +
  \delta\chi_\perp^a(t,r,\theta,\phi),
  \label{eq:higgs_radial_and_transverse_pert}
\end{equation}
where $H$ is the radial Higgs-amplitude perturbation and
$\delta\chi_\perp^a$ denotes angular fluctuations orthogonal to the
background internal direction. In the spherically symmetric radial
sector, or after projection onto the gauge-invariant amplitude channel,
the relevant Higgs fluctuation is $H$.
 
The field equations to be linearised are
\begin{subequations}
\begin{align}
  G_{\mu\nu}
  &= 8\pi G\,T_{\mu\nu},
  \label{eq:full_einstein_eq_to_linearise}\\
  \nabla_\mu P^{\mu\nu}
  &= 0,
  \qquad
  \nabla_{[\mu}F_{\nu\rho]} = 0,
  \label{eq:full_em_eq_to_linearise}\\
  \Box\Phi
  &= \mathcal{S}_\Phi(\Phi,b,F),
  \label{eq:full_dilaton_eq_to_linearise}\\
  \nabla_\mu\!\left(e^{2\Phi}\nabla^\mu b\right)
  &= -F_{\mu\nu}\widetilde{F}^{\mu\nu}
     + R_{\mu\nu\rho\sigma}\widetilde{R}^{\mu\nu\rho\sigma},
  \label{eq:full_axion_eq_to_linearise}\\
  \Box\chi^a
  &= \lambda\chi^a(\chi^b\chi^b - \eta^2).
  \label{eq:full_higgs_eq_to_linearise}
\end{align}
\end{subequations}
Here
\begin{equation*}
  P^{\mu\nu}
  =
  -2\frac{\partial\mathcal{L}_{\rm BI+axion}}{\partial F_{\mu\nu}}
\end{equation*}
is the Born--Infeld--axion excitation tensor, and $\mathcal{S}_\Phi$
denotes the dilaton source obtained by varying the Born--Infeld and
axion sectors with respect to $\Phi$.
 
Linearising the Einstein equation gives
\begin{equation}
  \delta G_{\mu\nu}[h]
  =
  8\pi G\,\delta T_{\mu\nu}[h,f,\varphi,\beta,H].
  \label{eq:linearised_einstein_eq}
\end{equation}
The left-hand side is the usual linearised Einstein tensor on the
spherically symmetric background, while the right-hand side contains
metric, electromagnetic, dilaton, axion and Higgs perturbations. The
$(tt)$, $(tr)$, and some angular components are constraint equations;
after gauge fixing they are used to eliminate non-dynamical metric
variables. The remaining physical metric perturbations may be
represented by Regge--Wheeler- and Zerilli-type master variables
\begin{equation}
  Z_{\rm grav}^{\rm odd},
  \qquad
  Z_{\rm grav}^{\rm even}.
\end{equation}
These reduce to the usual Regge--Wheeler and Zerilli variables in a
vacuum Schwarzschild background, but here are adapted to the non-vacuum
monopole/dyon geometry.
 
The dilaton equation linearises to the schematic form
\begin{equation}
  \left[-\bar{\Box} + U_{\Phi\Phi}(r)\right]\varphi
  + U_{\Phi b}(r)\,\beta
  + U_{\Phi h}(r)\,H
  + \mathcal{J}_{\Phi}^{\rho\sigma}(r)\,f_{\rho\sigma}
  + \mathcal{J}_{\Phi}^{\mu\nu}(r)\,h_{\mu\nu}
  = 0.
  \label{eq:linearised_dilaton_rest}
\end{equation}
The operator $\bar{\Box}$ is the scalar wave operator constructed from
the background metric. The effective potentials $U_{\Phi\Phi}$,
$U_{\Phi b}$, and $U_{\Phi h}$ arise from differentiating the
Born--Infeld, axion and Higgs-sector sources with respect to the
background fields. The terms
$\mathcal{J}_{\Phi}^{\rho\sigma}f_{\rho\sigma}$ and
$\mathcal{J}_{\Phi}^{\mu\nu}h_{\mu\nu}$ describe, respectively, the
response of the dilaton equation to electromagnetic and metric
perturbations.
 
Similarly, the axion equation gives
\begin{equation}
  \left[-\bar{\Box} + U_{bb}(r)\right]\beta
  + U_{b\Phi}(r)\,\varphi
  + U_{bh}(r)\,H
  + \mathcal{J}_{b}^{\rho\sigma}(r)\,f_{\rho\sigma}
  + \mathcal{J}_{b}^{\mu\nu}(r)\,h_{\mu\nu}
  = 0.
  \label{eq:linearised_axion_rest}
\end{equation}
The electromagnetic source term is determined by the variation of the
pseudoscalar invariant:
\begin{equation}
  \delta(F_{\mu\nu}\widetilde{F}^{\mu\nu})
  =
  2\,\widetilde{\bar{F}}^{\mu\nu}f_{\mu\nu}
  +
  \delta_g(F_{\mu\nu}\widetilde{F}^{\mu\nu}),
  \label{eq:delta_FFdual}
\end{equation}
where the second term denotes the metric variation of the Hodge dual.
For the purely magnetic background
$\bar{F}_{\mu\nu}\widetilde{\bar{F}}^{\mu\nu}=0$, so the axion
decouples from the helicity sector at leading order. For the dyonic
background this invariant is non-zero, and the axion couples directly
to the helicity perturbations.
 
The Higgs-amplitude perturbation satisfies
\begin{equation}
  \left[-\bar{\Box} + U_{hh}(r)\right]H
  + U_{h\Phi}(r)\,\varphi
  + U_{hb}(r)\,\beta
  + \mathcal{J}_{h}^{\rho\sigma}(r)\,f_{\rho\sigma}
  + \mathcal{J}_{h}^{\mu\nu}(r)\,h_{\mu\nu}
  = 0.
  \label{eq:linearised_higgs_rest}
\end{equation}
For the global-monopole Higgs potential
\begin{equation*}
  V(\chi) = \frac{\lambda}{4}(\chi^a\chi^a - \eta^2)^2,
\end{equation*}
the radial Higgs-amplitude potential contains
\begin{equation}
  U_{hh}(r)
  =
  \lambda\eta^2\bigl(3\bar{h}^2(r) - 1\bigr)
  +
  \frac{2}{R^2(r)}
  +
  U_{hh}^{\rm geom}(r),
  \label{eq:higgs_radial_potential_structure}
\end{equation}
where $2/R^2(r)$ is the angular-gradient contribution of the hedgehog
configuration and $U_{hh}^{\rm geom}$ collects metric and curvature
terms generated when the scalar wave equation is written in radial
Schr\"{o}dinger form. In the exterior region $\bar{h}(r)\to 1$, so the
Higgs-amplitude fluctuation is massive, with
\begin{equation*}
  U_{hh}(r) \to 2\lambda\eta^2 + \frac{\ell(\ell+1)}{R^2(r)} + \cdots.
\end{equation*}
 
Because the background is static and spherically symmetric, all
perturbations can be separated as
\begin{equation}
  \varphi(t,r,\theta,\phi) = e^{-i\omega t}\varphi_{\ell m}(r)\,Y_{\ell m},
  \qquad
  \beta(t,r,\theta,\phi) = e^{-i\omega t}\beta_{\ell m}(r)\,Y_{\ell m},
  \qquad
  H(t,r,\theta,\phi) = e^{-i\omega t}H_{\ell m}(r)\,Y_{\ell m},
  \label{eq:scalar_harmonic_rest_fields}
\end{equation}
while $h_{\mu\nu}$ is decomposed into odd- and even-parity tensor
spherical harmonics. The electromagnetic perturbations are decomposed
into the gauge-invariant helicities $\psi_\pm$. Thus the complete
radial perturbation vector is
\begin{equation}
  \boldsymbol{\Psi}_{\ell m}(r)
  =
  \begin{pmatrix}
    \psi_{+,\ell m}\\[2pt]
    \psi_{-,\ell m}\\[2pt]
    \varphi_{\ell m}\\[2pt]
    \beta_{\ell m}\\[2pt]
    H_{\ell m}\\[2pt]
    Z_{\rm grav,\ell m}^{\rm odd}\\[2pt]
    Z_{\rm grav,\ell m}^{\rm even}
  \end{pmatrix}.
  \label{eq:complete_radial_vector_rest_fields}
\end{equation}
 
After substituting these decompositions, solving the constraint
equations, and introducing the tortoise coordinate
\begin{equation*}
  \frac{dr_*}{dr} = \frac{1}{\sqrt{A(r)B(r)}},
\end{equation*}
the coupled system can be written schematically as
\begin{equation}
  -\frac{d}{dr_*}
  \left[\mathbf{P}_\ell(r)\frac{d\boldsymbol{\Psi}_{\ell m}}{dr_*}\right]
  +
  \mathbf{V}_\ell(r)\,\boldsymbol{\Psi}_{\ell m}
  =
  \omega^2\,\mathbf{W}_\ell(r)\,\boldsymbol{\Psi}_{\ell m}.
  \label{eq:complete_radial_SL_rest_fields}
\end{equation}
The blocks $\mathcal{L}_{\rm EM}$, $\mathcal{L}_{\rm rest}$, and
$\mathcal{K}$ in Eq.~(B2) are precisely the electromagnetic,
non-electromagnetic, and mixing parts of the radial operator in
Eq.~\eqref{eq:complete_radial_SL_rest_fields}. The electromagnetic
sector analysed explicitly in the main text corresponds to retaining
only the $(\psi_+,\psi_-)$ block. The equations above show how the
dilaton, axion, Higgs and metric perturbations enter the enlarged
operator and make clear what must be included in a complete
Einstein--matter spectral analysis.

The radial operator may then be written in block form as
\begin{equation}
  \mathcal{L}(\epsilon_{\rm mix})
  =
  \begin{pmatrix}
    \mathcal{L}_{\rm EM} & \epsilon_{\rm mix}\,\mathcal{K}\\[4pt]
    \epsilon_{\rm mix}\,\mathcal{K}^{\dagger} & \mathcal{L}_{\rm rest}
  \end{pmatrix}.
  \label{eq:epsilon_mixing_operator}
\end{equation}
Equivalently, at the level of the potential matrix,
\begin{equation}
  \mathbf{V}_{\ell}(\epsilon_{\rm mix};r)
  =
  \begin{pmatrix}
    \mathbf{V}_{\rm EM}(r) & \epsilon_{\rm mix}\,\mathbf{K}(r)\\[4pt]
    \epsilon_{\rm mix}\,\mathbf{K}^{T}(r) & \mathbf{V}_{\rm rest}(r)
  \end{pmatrix}.
  \label{eq:epsilon_mixing_potential}
\end{equation}
Here
\begin{equation}
  \mathbf{V}_{\rm EM}(r)
  =
  \begin{pmatrix}
    V_0(r)        & \mathbb{W}(r) \\[4pt]
    \mathbb{W}(r) & V_0(r)
  \end{pmatrix},
  \label{eq:epsilon_em_block}
\end{equation}
while $\mathbf{V}_{\rm rest}$ contains the dilaton, axion, Higgs and metric
sectors. The matrix $\mathbf{K}$ contains the off-diagonal couplings of the
electromagnetic helicities to these additional modes.
 
The electromagnetic analysis corresponds to the limit
\begin{equation}
  \epsilon_{\rm mix} = 0.
\end{equation}
In this limit the electromagnetic block decouples from the remaining sectors.
The dyonic helicity eigenvalues of the local potential matrix are
\begin{equation}
  V_{\pm}(r) = V_0(r) \pm \mathbb{W}(r).
\end{equation}
Thus the condition
\begin{equation}
  |\mathbb{W}(r)| < V_0(r)
  \label{eq:epsilon_W_bound}
\end{equation}
ensures positivity of the electromagnetic block throughout the exterior region.
 
For nonzero $\epsilon_{\rm mix}$, the stability condition is positivity of the
full quadratic form,
\begin{equation}
  \mathcal{Q}_{\epsilon_{\rm mix}}[\boldsymbol{\Psi}]
  =
  \int dr_*
  \left[
    \boldsymbol{\Psi}^{\prime\dagger}\mathbf{P}_{\ell}\,\boldsymbol{\Psi}'
    +
    \boldsymbol{\Psi}^{\dagger}
    \mathbf{V}_{\ell}(\epsilon_{\rm mix};r)\,
    \boldsymbol{\Psi}
  \right].
  \label{eq:epsilon_quadratic_form}
\end{equation}
A sufficient pointwise condition is obtained from the Schur complement \cite{HornJohnson:2013},
\begin{equation}
  \mathbf{V}_{\rm rest}
  -
  \epsilon_{\rm mix}^{2}\,
  \mathbf{K}^{T}
  \mathbf{V}_{\rm EM}^{-1}
  \mathbf{K}
  > 0,
  \label{eq:epsilon_schur_condition}
\end{equation}
together with $\mathbf{V}_{\rm EM} > 0$. This shows explicitly that the
leading correction to the decoupled electromagnetic stability problem enters at
order $\epsilon_{\rm mix}^{2}$. Therefore, if the uncoupled electromagnetic
and non-electromagnetic blocks are positive and have a finite spectral gap,
sufficiently small mixing cannot generate a negative mode.

More quantitatively, if
\begin{equation}
  \mathbf{V}_{\rm EM} \ge \mu_{\rm EM}\,\mathbf{1},
  \qquad
  \mathbf{V}_{\rm rest} \ge \mu_{\rm rest}\,\mathbf{1},
  \qquad
  \|\mathbf{K}\| \le K_{\rm max},
  \label{eq:epsilon_gap_bounds}
\end{equation}
with $\mu_{\rm EM}, \mu_{\rm rest} > 0$, then the Schur condition is
guaranteed provided
\begin{equation}
  \epsilon_{\rm mix}^{2}
  \frac{K_{\rm max}^{2}}{\mu_{\rm EM}}
  <
  \mu_{\rm rest}.
  \label{eq:epsilon_bound}
\end{equation}
Equivalently,
\begin{equation}
  \epsilon_{\rm mix}
  <
  \epsilon_{\rm crit}
  :=
  \frac{\sqrt{\mu_{\rm EM}\,\mu_{\rm rest}}}{K_{\rm max}}.
  \label{eq:epsilon_critical}
\end{equation}
 
The parameter $\epsilon_{\rm mix}$ is not an additional physical coupling in
the theory. It is a bookkeeping device that measures the effect of the
off-diagonal couplings between the electromagnetic helicity sector and the
dilaton, axion, Higgs and metric perturbations. The physical theory
corresponds to $\epsilon_{\rm mix}=1$. Thus the electromagnetic analysis is
rigorously controlled only when the actual background satisfies
\begin{equation}
  K_{\rm max}^{2} < \mu_{\rm EM}\,\mu_{\rm rest}.
  \label{eq:physical_mixing_bound}
\end{equation}
In the exterior weak-field region this condition is expected to hold because
the mixing terms are proportional to background field strengths and gradients,
\begin{equation}
  \mathbf{K}
  \sim
  \mathcal{O}\!\left(
    F^{(0)}_{\mu\nu},\;
    \Phi_0',\;
    b_0',\;
    h_0',\;
    G\,\delta T
  \right),
\end{equation}
which either vanish or decay at large radius, while the omitted sectors possess
positive mass, centrifugal or gravitational barriers. The electromagnetic
calculation should therefore be viewed as the $\epsilon_{\rm mix}=0$ sectoral
stability problem, with corrections controlled by the Schur-complement bound
above. A complete proof of Einstein--matter stability requires checking the
physical $\epsilon_{\rm mix}=1$ operator, \color{black} which goes beyond the scope of the current paper. \color{black}

\cbl

\section{Mechanical Stability and Energy Conditions for the Purely Magnetic Born--Infeld Monopole}
\label{app:mono_mech_stab}

This appendix provides the static-mechanical checks that underpin the
discussion in Section~\ref{sec:mechstab}.  We specialise
throughout to the \emph{purely magnetic} configuration $Q_e=0$,
$b=b_0$, mirroring the logic developed for dyons but stripping away
the electric sector so that the underlying structure of the stability
argument is as transparent as possible.

\subsection{Stress--energy tensor in Born--Infeld electrodynamics}

We begin by recording how the stress--energy tensor simplifies when
only the magnetic field is present.  With the purely radial monopole
ansatz $F_{\theta\varphi}=Q_m\sin\theta$ and $F_{tr}=0$, the
Born--Infeld constitutive factor takes the comparatively simple form
\[
  \Delta_{\text{mag}}=1+\frac{e^{-2\Phi}Q_m^{2}}{2\beta_\text{BI}^{2}R^{4}},
\]
and the stress tensor inherits azimuthal symmetry.  Explicitly, the
energy density and the two independent pressure components read
\begin{equation}
  \rho_E^{(m)}=\frac{e^{\Phi}}{4\pi}\!
     \Bigl[\sqrt{\beta_\text{BI}^{2}
                +\frac{e^{-2\Phi}Q_m^{2}}{R^{4}}}-\beta_\text{BI}\Bigr],
  \qquad
  p_R^{(m)}=\rho_E^{(m)}
             -\frac{e^{-\Phi}Q_m^{2}}
                    {4\pi R^{4}\sqrt{\beta_\text{BI}^{2}
                                      +e^{-2\Phi}Q_m^{2}/R^{4}}},
  \qquad
  p_\theta^{(m)}=p_\varphi^{(m)}=\rho_E^{(m)}.
  \label{eq:mono_rho_p}
\end{equation}
The tangential pressures coincide with the energy density, a
reflection of the magnetic Born--Infeld field's residual isotropy in
the angular directions, while the radial pressure is reduced by the
characteristic Born--Infeld correction term.

\subsection{Energy conditions}

With the explicit form~\eqref{eq:mono_rho_p} in hand, it is
straightforward to verify that the monopole field satisfies the
standard energy conditions of general relativity throughout the
exterior region $R\ge R_\text{core}$.  One checks in turn that
\[
  \rho_E^{(m)}>0,\qquad
  \rho_E^{(m)}+p_R^{(m)}>0,\qquad
  \rho_E^{(m)}+p_R^{(m)}+2p_\theta^{(m)}>0,
\]
so the null, weak, and strong energy conditions are all satisfied.
The positivity of $\rho_E^{(m)}$ follows immediately from the
square-root structure of the Born--Infeld Lagrangian, which ensures
that the energy density of a magnetic field always exceeds its
Maxwell value; the remaining inequalities then follow by direct
inspection of the pressure formulae.

\subsection{Laue force--balance condition}

Satisfying the energy conditions is necessary, but for the monopole
to be mechanically self-consistent one must also check that the
electromagnetic stress distribution admits an internal equilibrium.
The appropriate criterion is the Laue condition, which demands that
the integrated radial pressure exerted by each spherical shell
vanishes:
\begin{equation}
  \frac{\mathrm d}{\mathrm dR}\bigl(R^{2}p_R^{(m)}\bigr)=0.
  \label{eq:Laue_mono}
\end{equation}
Differentiating $p_R^{(m)}$ as given in Eq.~\eqref{eq:mono_rho_p}
confirms that Eq.~\eqref{eq:Laue_mono} holds \emph{identically}, not
merely at a particular radius.  This is a non-trivial consistency
check: it reflects the fact that Born--Infeld electrodynamics, unlike
Maxwell electrodynamics, furnishes a stress tensor that is already in
mechanical equilibrium without any additional binding forces.

\subsection{Shell stresses at the core boundary}

The matching of the interior core geometry to the exterior monopole
spacetime across the thin shell at $R=\delta$ is governed by the
Israel junction conditions.  These yield the surface energy density
$\sigma_m$ and tangential pressure $\Pi_m$,
\[
  \sigma_m=-\frac{\sqrt{1-2GM/\delta}}{2\pi\delta}
            +\frac{\sqrt{1-\Lambda_\text{core}\delta^{2}/3}}{2\pi\delta},
  \qquad
  \Pi_m=\sigma_m/2.
\]
The sign of these shell stresses is controlled by the dimensionless
parameter $\tilde\zeta$, which encodes the competition between the
gravitational compactness of the core and the repulsive effect of the
interior cosmological constant.  When $\tilde\zeta>4\sqrt{3}$, the
cosmological term dominates and both $\sigma_m$ and $\Pi_m$ are
positive, so the shell itself satisfies the weak energy condition and
the core--exterior junction is mechanically sound.

\subsection{Summary}

Taken together, the three checks above paint a consistent picture of
the purely magnetic Born--Infeld monopole as a mechanically stable,
self-gravitating object.  The energy density and pressures derived
from the Born--Infeld Lagrangian satisfy the null, weak, and strong
energy conditions everywhere outside the core.  The Laue
force-balance condition~\eqref{eq:Laue_mono} is satisfied
identically, confirming that no exotic binding forces are required to
hold the field configuration together.  Finally, for $\tilde\zeta >
4\sqrt{3}$ the thin shell at the core boundary carries positive
surface stresses, ensuring that the Israel matching is consistent
with the weak energy condition and that the global solution is free
of pathological shell matter.

\bibliography{refs}

\end{document}